\documentclass[journal]{IEEEtran}

\usepackage{color}
\usepackage{graphicx}
\usepackage{cite}
\usepackage{amsmath}
\usepackage{booktabs}
\usepackage{multirow}
\usepackage{array}
\usepackage{nccmath}
\usepackage{textcomp}
\usepackage{textgreek}
\usepackage{boldline} 
\usepackage{lipsum}
\usepackage{epstopdf}
\usepackage{tabularx}
\usepackage{babel}
\usepackage{tikz}
\usepackage{stackengine}
\usepackage{titlesec}
\usepackage[normalem]{ulem}
\usepackage[dvips]{epsfig}
\usepackage[LGRgreek]{mathastext}

\ifCLASSINFOpdf
 
\else

\fi

\AtBeginDocument{\let\latexlabel\label}
\begin{document}

\title{DustNet: A Wireless Network of Ultrasonic Neural Implants}

\author{Jade~Pinkenburg$^{*1}$,~\IEEEmembership{Graduate Student Member,~IEEE,}
    Changuk~Lee$^{*1}$,~\IEEEmembership{ Member,~IEEE,} Mohammad~Meraj~Ghanbari$^{*1}$,~\IEEEmembership{Member,~IEEE,}
    Cem~Yalcin$^1$,~\IEEEmembership{Member,~IEEE,}
    Miguel~Montalban$^1$,
    and~Rikky~Muller$^{1,2}$,~\IEEEmembership{Senior Member,~IEEE}%

\thanks{This work was supported in part by the National Science Foundation (NSF) Graduate Research Fellowship Program under Grant No. DGE 2146752 and in part by the Weill Neurohub Fellowship.}
\thanks{
C. Lee, J. Pinkenburg, M.M. Ghanbari, C. Yalcin, and M. Montalban are with the Department of Electrical Engineering and Computer Sciences, University of California at Berkeley, Berkeley, CA, 94720 USA.}
\thanks{
R. Muller is with the Department of Electrical Engineering and Computer Sciences, University of California at Berkeley, Berkeley, CA, 94720 USA and also with Weill Neurohub, Berkeley, CA, 94720 USA.
}
}

\markboth{IEEE Transactions on Biomedical Circuits and Systems}
{Pinkenburg \MakeLowercase{\textit{et al.}}: }

\maketitle

\begin{abstract}

Spatially distributed peripheral nerve recordings can be used to reconstruct motor intention and improve natural control of prosthetics.  However, many existing clinical solutions rely on percutaneous wires to access peripheral nerves; these sites are prone to infection and motion-induced electrode degradation, preventing chronic use. To address the need for fully wireless neural recording systems, this paper presents DustNet: a spatially-distributed network of ultrasonically-powered neural recording implants capable of supporting up to 8 simultaneously-recording nodes over a single ultrasound link. To enable high-throughput multi-implant communication, DustNet implements a time-division multiple-access (TDMA) protocol with up to 16-level amplitude modulation of the ultrasound backscatter that achieves up to 4x higher data rates than traditional on-off keying methods. Each neural implant consists of a 0.7x0.7x0.7 $\text{mm}^\text{3}$ piezoceramic transducer, a 100 nF off-chip capacitor, and an IC mounted on a flexible PCB. The implant IC was fabricated in a 28nm CMOS process and occupies an area of 0.43 $\textbf{mm}^{\textbf{2}}$. System functionality was verified at 90mm depth in oil, achieving a maximum measured data rate of 200 kb/s at 2 MHz ultrasound carrier frequency, with each implant transmitting uplink data at 50 kb/s and dissipating just 7 µW; the system is demonstrated to support up to 400 kb/s total data rate over the same link.

\end{abstract}


\section{Introduction}

\IEEEPARstart{P}ERIPHERAL nerves are responsible for carrying motor and sensory signals throughout the body; developing implantable devices to record activity from these nerves has myriad applications in basic neuroscience, disease diagnosis, and therapeutic treatments \cite{lee_implantable_2018}. For example, recordings of efferent motor nerve signals from functional nerve endings retained by patients with limb deficiencies can be used to decode motion intention. Recent studies have demonstrated that real-time activity data recorded from these nerves can be used to significantly improve the performance of prosthetic limbs and give patients more intuitive and reliable control over these devices \cite{Vu2020}. However, existing clinical efforts have primarily relied on percutaneous wires to connect external circuitry with implanted neural electrodes. These sites tend to be very susceptible to infection. External forces on the wires can also cause signal quality to degrade over time by shifting the electrode placement. As a result, the percutaneous electrodes must be explanted after several weeks, preventing long-term use \cite{Vancara23}. 

\begin{figure}[t]
\centering
\includegraphics[width=8.0Cm]{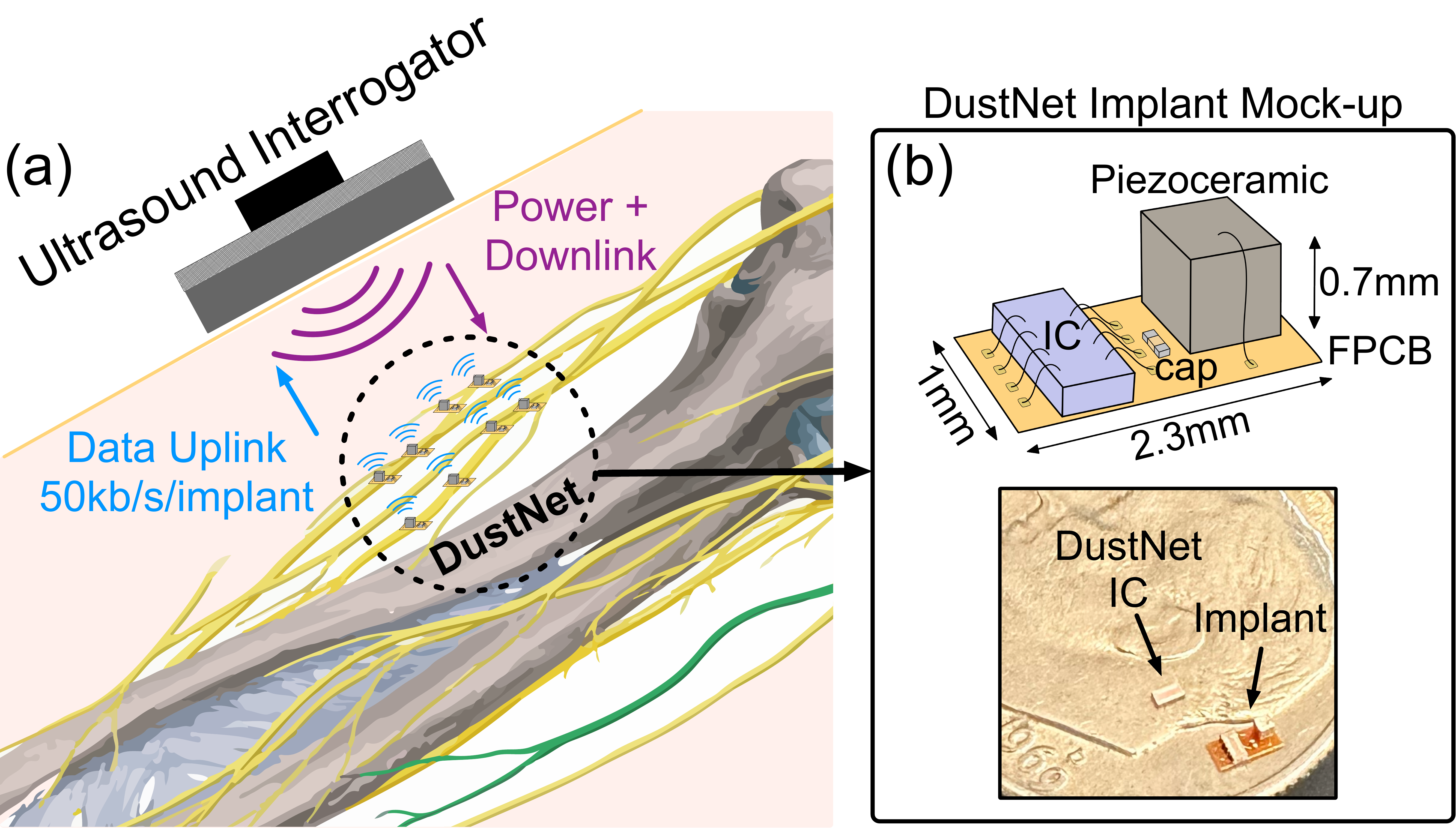}
\setlength\abovecaptionskip{1pt}
\caption{(a) Conceptual diagram of the proposed DustNet wireless neural recording platform with 8 implanted sensor nodes. (b) Implant mock-up diagram and assembly with a US dime for reference.}
\label{fig_concept}
\vspace{-8pt}
\end{figure}

In contrast, miniaturized neural recording implants with wireless power transfer and data transmission enable chronic recording of peripheral nerve activity with minimal invasiveness to the patient. Because several distinct nerves coordinate limb movement, a spatially-distributed network of simultaneously recording implants is required to ensure dexterous control of limb prosthetics. Previous works have demonstrated networks of multiple neural recording implants using radiofrequency (RF), near-infrared (NIR), magnetoelectric (ME), or ultrasound (US) links \cite{lee_neural_2021, Atzeni20, Yu2020,Ghanbari_JSSC19,alamouti_high_2020}. Although RF carrier frequencies enable high overall data rates, the high attenuation coefficient of RF energy in tissue severely limits the depth of these implants \cite{lee_neural_2021}, preventing recording from deeply-seated peripheral motor nerves; NIR similarly suffers from significant attenuation in tissue \cite{Atzeni20}. Conversely, magnetoelectric links penetrate deep into tissue with low loss, but suffer from limited per-implant data rates (less than 10 kb/s/implant) \cite{Yu2020}. In contrast, US links enable low-loss communication in the body while offering sufficient channel bandwidth to support transmission of continuous neural activity data at depth \cite{Ghanbari_JSSC19, alamouti_high_2020}. Because the transducer size often dominates the overall wireless implant volume, US links also enable mm-scale implant miniaturization due to the sub-mm wavelength of US in the body at frequencies of interest \cite{Ghanbari_TBIOCAS20}. The safety limit for transmitted US power in the body is also significantly higher than that for RF energy, resulting in higher received power at the implant and further enabling transducer miniaturization. However, increasing tissue absorption at higher frequencies limits US carrier frequencies to the low-MHz range in the human body. 

High-fidelity spike sorting requires a recording bandwidth of at least 3 kHz \cite{rey_past_2015}, but robust decoding of muscle intention can also be achieved using low frequency local field potential (LFP) or electromyography (EMG) envelopes that require lower recording bandwidths \cite{flint2012local, stavisky2015high}. Prior clinical studies have demonstrated that recording from 8 distinct nerve fascicles is sufficient to achieve near-natural control over limb prosthetics \cite{Vu2020}. Thus, a total system data rate of at least 400 kb/s is required to ensure continuous and intuitive prosthetic control. To enable recording from deeply-seated motor nerves (e.g., sciatic nerve for leg prosthetics), the implants must also be able to operate at depths of over 50mm. Although US links have been demonstrated to power implants at this depth in tissue, the low US carrier frequency necessitates a spectrally-efficient communication scheme to achieve the required system data rate. To address these challenges, we present DustNet: a network of wireless, ultrasonically-powered neural implants capable of supporting up to 8 simultaneously recording implants using a single US link for power transfer and bidirectional data communication. DustNet implements a custom time-division multiple-access (TDMA) protocol to communicate with multiple implants while minimizing inter-implant interference. To achieve a high data rate despite the low carrier frequency of ultrasound in the human body, we introduce an efficient 16-level amplitude shift keying (ASK) modulation scheme to encode multiple bits per symbol in the US backscatter. Using these techniques, DustNet supports an overall data transmission rate of 400 kb/s at 2 MHz US carrier frequency, with a per-implant data rate of 50 kb/s. Each implant occupies a volume smaller than 0.5 $mm^3$ and consumes 7 µW. System operation is verified at 90mm depth in oil and the bit-error rate (BER) of the communication protocol is measured to be at most 1.89E-5.

This paper is an extension of \cite{CLee_ISSCC25}, and offers an extended discussion of the TDMA communication protocol, analysis of the multi-level ASK backscatter scheme, an updated circuit implementation including an analog front-end for recording neural signals, and presents additional measurement results. Section II discusses the communication protocol, Section III explains the chip implementation, and section IV reports the measurement results. Section V presents a comparison against state-of-the-art neural implants and summarizes this work.

\section{Communication Protocol}

\subsection{System Overview}

Fig. 1 illustrates the proposed wireless neural recording system. DustNet consists of an unfocused ultrasound transducer external to the body and up to 8 free-floating neural recording devices to be implanted on peripheral nerves. The external transducer is placed in contact with the skin and transmits US pulses through tissue; the implants each harvest power using a miniaturized piezoceramic transducer (piezo) to convert the mechanical US wave into an electrical AC signal that is rectified to DC power by the implant IC and stored on an off-chip capacitor. To transmit data, the implants utilize a power-efficient pulse-echo communication scheme in which the implant encodes data in the backscattered acoustic wave. Because DustNet is specifically intended to be used for peripheral nerve recording, the acoustic propagation path from the skin-mounted external interrogator solely passes through soft tissues with similar acoustic impedances. This avoids the signal degradation, reflection, and scattering caused by high-impedance bone-tissue interfaces that can degrade the performance of US-based communication protocols.

\subsection{Multiple-Access Ultrasound Communication}

\begin{figure}[t]
\centering
\includegraphics[width=8.0Cm]{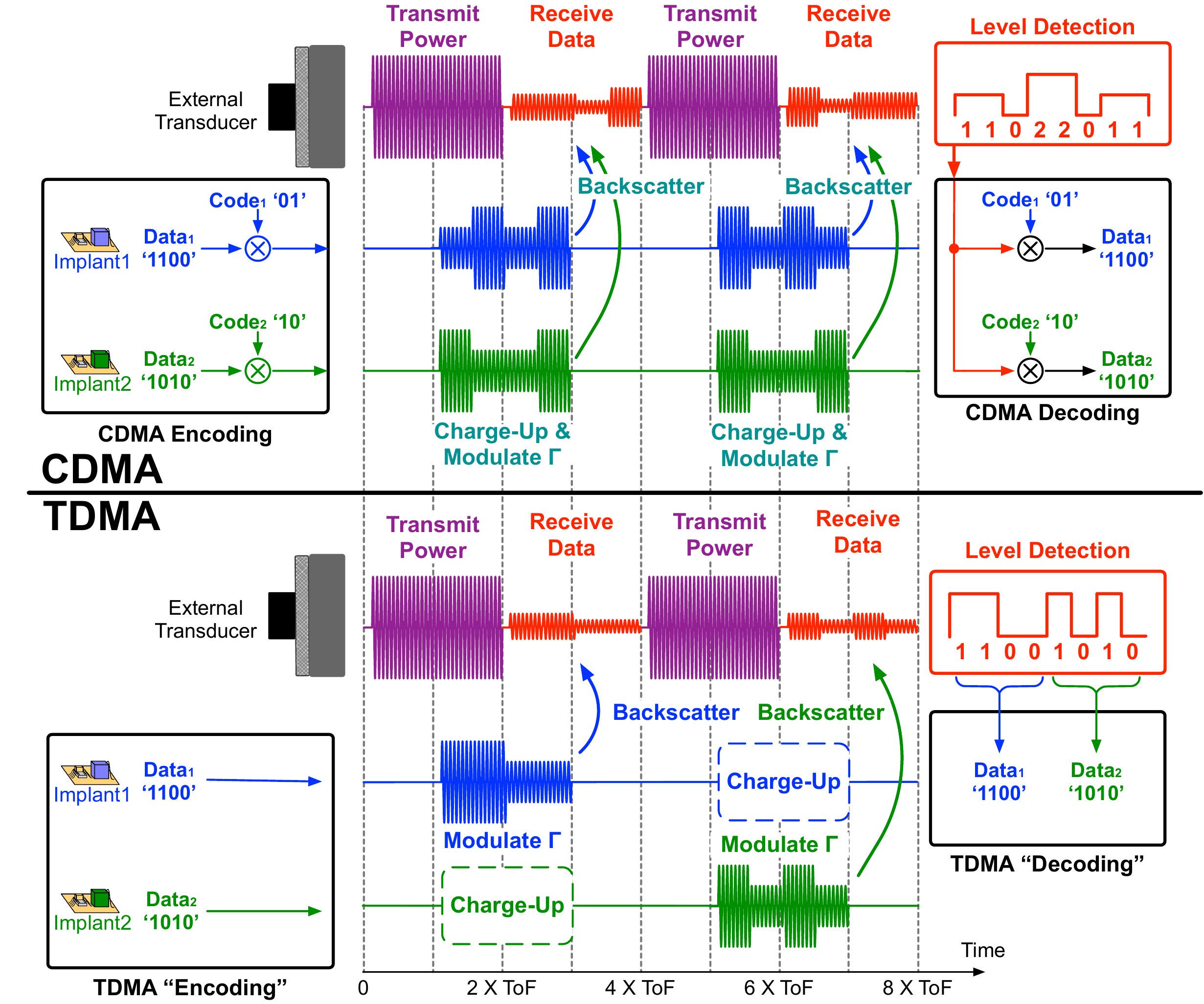}
\setlength\abovecaptionskip{1pt}
\caption{Timing diagram for pulse-echo communication using CDMA and TDMA protocols. During uplink, the maximum achievable data rate for both protocols is identical, but TDMA is more robust to inter-implant mismatch and allows implants to charge without sacrificing channel bandwidth.}
\label{fig_CDMA_vs_TDMA}
\end{figure}

Previous work has demonstrated full-duplex ultrasound communication with continuous power transfer by using one piezo for downlink power and data while actively driving a second piezo for uplink communication \cite{Chang17}. However, the large power consumption demanded by the active drive circuitry limits the implant depth and the second piezo significantly increases the implant volume. To minimize implant size and power consumption, DustNet employs a single-piezo, pulse-echo communication scheme that eliminates the need for active drive, as demonstrated in \cite{Ghanbari_JSSC19, Seo16}. Although this precludes continuous power or data transmission, the pulse-echo scheme minimizes implant power requirements and simplifies requirements on the external transceiver by eliminating the need for full-duplex hardware or multiple transducers. Lossless data transmission and continuous implant power are maintained by on-device charge storage and data memory. To transmit data, each implant modulates the acoustic reflection coefficient (Γ) of its piezo to encode data on the backscattered acoustic wave. Simultaneous bidirectional transmission is avoided due to the high dynamic range required to operate the transducer in both RX and TX mode; TX mode requires high-voltage (50V\textsubscript{pp}), and the reflected waves generate received signal voltages of just a few mV\textsubscript{pp} in RX mode.

To enable communication with multiple implants, previous works have employed cellular techniques such as TDMA \cite{Chang19}, code-division multiple access (CDMA) \cite{Ghanbari_JSSC19, alamouti_high_2020}, and frequency-division multiple access (FDMA) \cite{Chang19} protocols for data transmission over an ultrasound link, all of which are compatible with a pulse-echo communication scheme. Because the implant piezos have a narrow bandwidth, FDMA is difficult to implement for large systems since it necessitates that the implant piezos are of precise, varying sizes to generate different resonant frequencies; this significantly increases overall system complexity, making CDMA and TDMA protocols much more practical choices. Assuming perfect use of the available channel bandwidth, the channel capacity of CDMA and TDMA communication links are identical for a fixed carrier frequency, as shown in Fig. 2. However, CDMA is particularly sensitive to variations in implant depth and mismatches in the signal power transmitted by each implant. For example, \cite{alamouti_high_2020} introduces an ISI-tolerant US CDMA decoder that supports high total data rates, but required a complex machine learning receiver architecture to correct for path differences and implant mismatch. In contrast, TDMA is robust to inter-implant interference because only one implant transmits during each uplink timeslot. Depth mismatch is addressed by adjusting the length of the uplink timeslot on a per-implant basis to account for variations in the time-of-flight to each implant.

To receive data from multiple implants, DustNet employs a custom pulse-echo TDMA protocol in which channel control cycles sequentially between implants. In this protocol, the external transducer continuously transmits power pulses to the implants and only one implant backscatters data on each pulse. The remaining implants harvest power from the US pulse and charge their storage capacitors. The uplink order is assigned during implant configuration, and the implants count the incoming US pulses to keep track of channel control. The external transducer and implants must be aligned such that each DustNet node receives sufficient US power and the backscattered wave is incident on the transducer surface. Previous studies have characterized US-based implant performance degradation due to lateral and angular misalignment \cite{Roschelle_JSSC, Seo16}. In DustNet, peak performance is achieved when the interrogator and implants are aligned on-axis, but the TDMA protocol and adaptive ASK thresholding scheme in the receive chain (see Section IV-A) helps mitigate performance degradation caused by lateral or angular displacements.

\begin{figure}[t]
\centering
\includegraphics[width=8.0Cm]{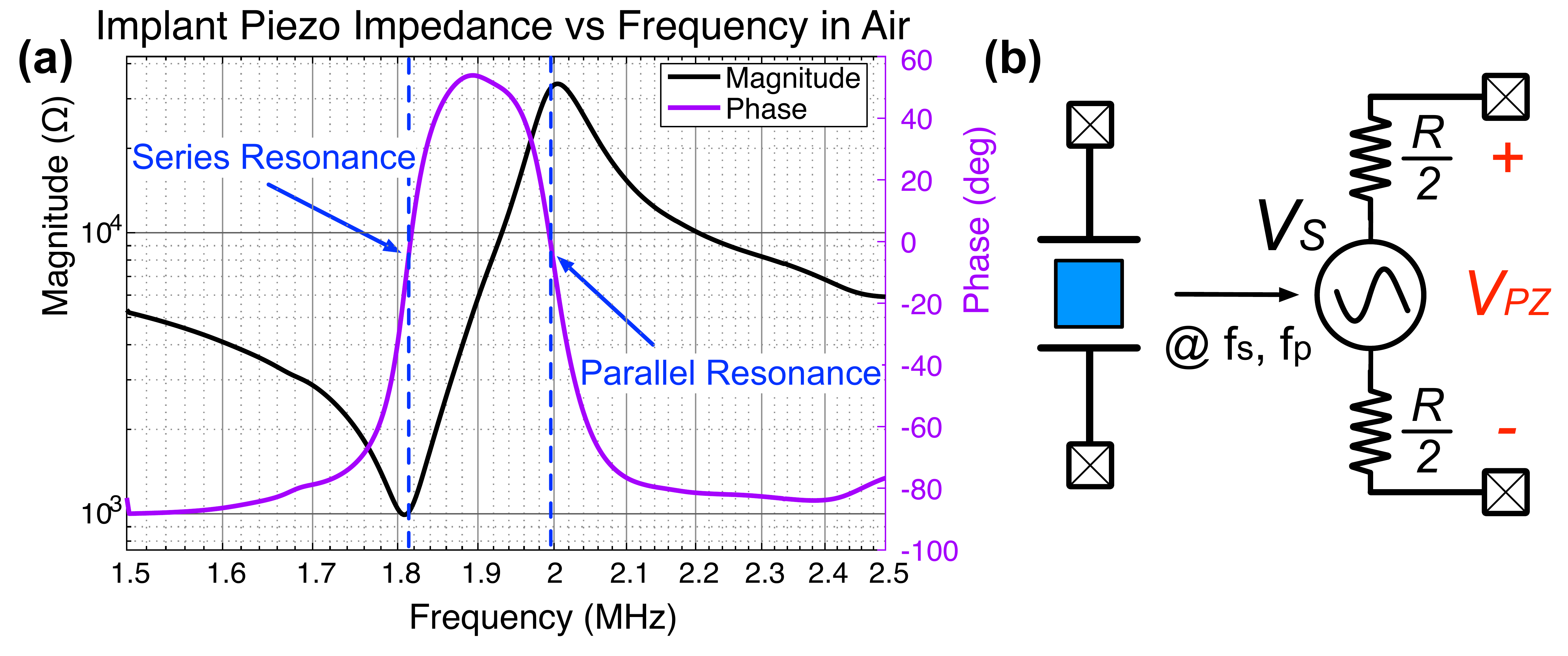}
\setlength\abovecaptionskip{1pt}
\caption{(a) Typical implant piezo impedance as a function of frequency. At resonance ($f_s$ and $f_p$), the piezo impedance is purely resistive. (b) Piezo electrical model at resonance. The resistance at the series resonance is significantly lower than the resistance at the parallel resonance.}
\label{fig_pz_measurements}
\end{figure}

\subsection{Multi-Level ASK Modulation}

Modulating the acoustic reflection coefficient by adjusting the piezo termination impedance $Z_E$ encodes data on the amplitude of the backscattered wave. However, the nonlinear relationship between Γ and $Z_E$ has caused significant distortion in previously developed analog backscatter modulators \cite{Seo16,Ozilgen17}. Furthermore, the nonlinear transfer function amplifies the SNR degradation due to electronic and carrier noise at compressed lower amplitudes. Discrete modulation schemes are inherently more robust to carrier noise, but transmitting multi-level discrete data over the nonlinear analog channel results in unevenly spaced amplitude levels; this reduces the minimum distance ($d_{min}$) between adjacent symbols and degrades the BER. Previous works \cite{Roschelle_JSSC, Sonmezoglu20} have implemented binary ASK backscatter modulation schemes for uplink data transmission to maximize $d_{min}$. Although this method is maximally robust to carrier noise and minimizes system complexity, the large spacing between echo levels is spectrally inefficient for low-noise channels. Furthermore, the overall channel capacity is fundamentally limited by the narrow bandwidth of the piezoelectric transducer; the resonance restricts the maximum allowable symbol rate before ISI significantly degrades the link. Therefore, to increase the data rate while maintaining a sufficiently low BER, a multi-level ASK modulation scheme can be employed to harness the trade-off between spectral efficiency and noise robustness.

To implement the linear backscatter modulator required for multi-level backscatter amplitude modulation, accurate modeling of the implant piezo is required. Recent work has demonstrated that at resonance, Γ is dependent on the electrical impedance across the piezo terminals and the internal impedance of the piezo \cite{Ghanbari_TBIOCAS20}. A typical impedance profile of the implant piezo as a function of frequency is shown in Fig. 3(a), and it exhibits two resonant frequencies: a series resonant frequency ($f_s$) at 1.82 MHz and a parallel resonant frequency ($f_p$) at 2.00 MHz. The resonant frequency of the transducer is measured using an impedance analyzer. The piezo impedance is much higher at the parallel resonant frequency than at the series resonant frequency. At resonance, the impedance of the piezo is purely resistive, and the transducer can be modeled as a Thevenin equivalent circuit without reactive components, as shown in Fig. 3(b). Equations (1) and (2) demonstrate how Γ is determined for the series and parallel resonant frequencies, where $Z_{Th}$ is the Thevenin equivalent resistance at resonance and $Z_E$ is the termination impedance across the piezo terminals.

\begin{equation*}
\refstepcounter{equation}\latexlabel{firsthalf}
\refstepcounter{equation}\latexlabel{secondhalf}
\Gamma_p \propto \frac{Z_{Th,s}}{Z_{E}+Z_{Th,s}} \qquad \Gamma_s \propto \frac{Z_{E}}{Z_{E}+Z_{Th,p}}
\tag{\ref{firsthalf}, \ref{secondhalf}}
\end{equation*}

\begin{equation*}
\refstepcounter{equation}\latexlabel{eq3}
\refstepcounter{equation}\latexlabel{eq4}
\Gamma_p \propto V_{S}-V_{PZ} \qquad \Gamma_s \propto V_{PZ}
\tag{\ref{eq3}, \ref{eq4}}
\end{equation*}

\begin{figure}[t]
\centering
\includegraphics[width=8.0Cm]{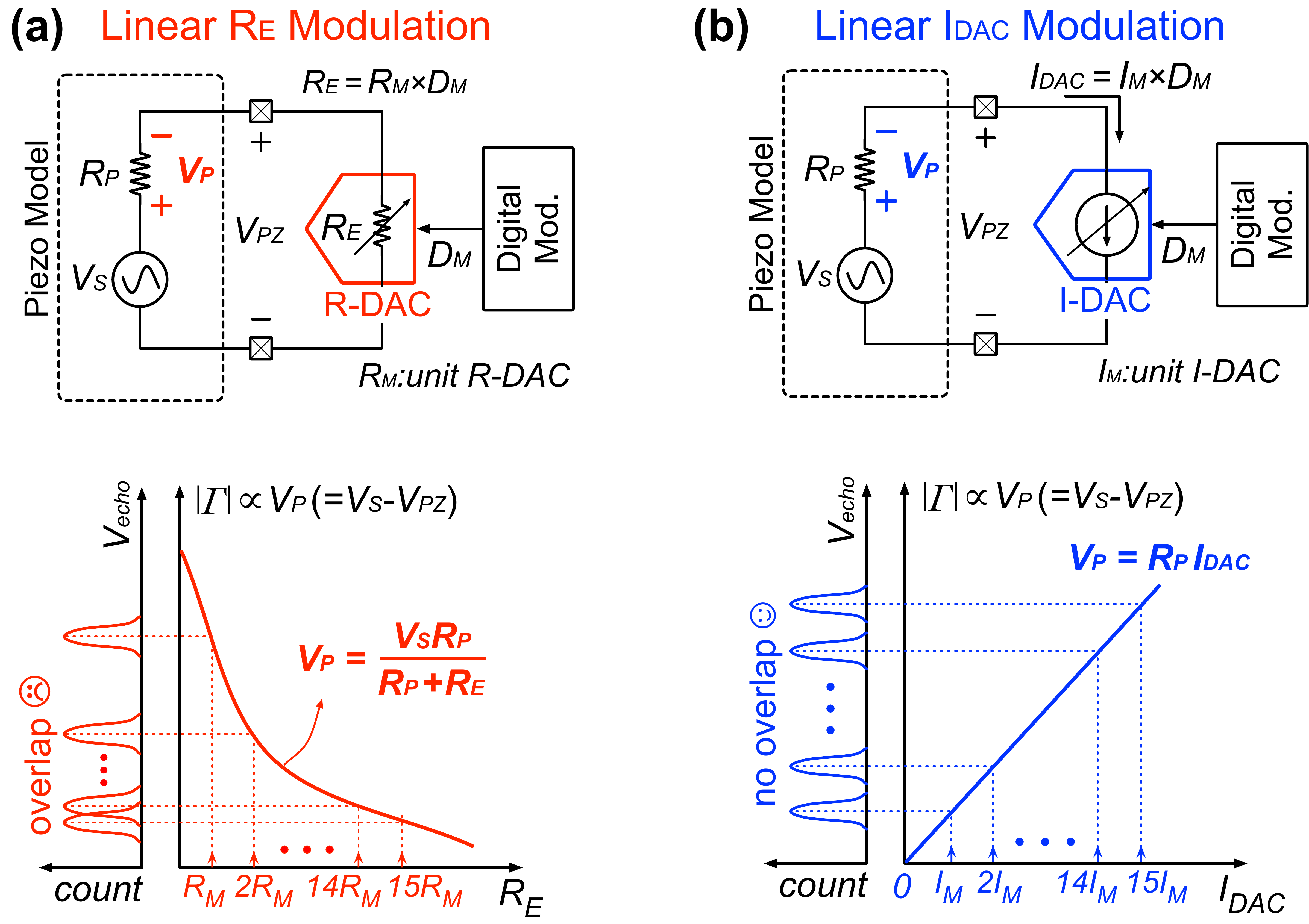}
\setlength\abovecaptionskip{1pt}
\caption{Multi-level ASK modulation: (a) Using a linear R-DAC and (b) using a linear I-DAC.}
\label{fig_mASK}
\end{figure}

\begin{figure*}[ht]
\centering
\includegraphics[width=17Cm]{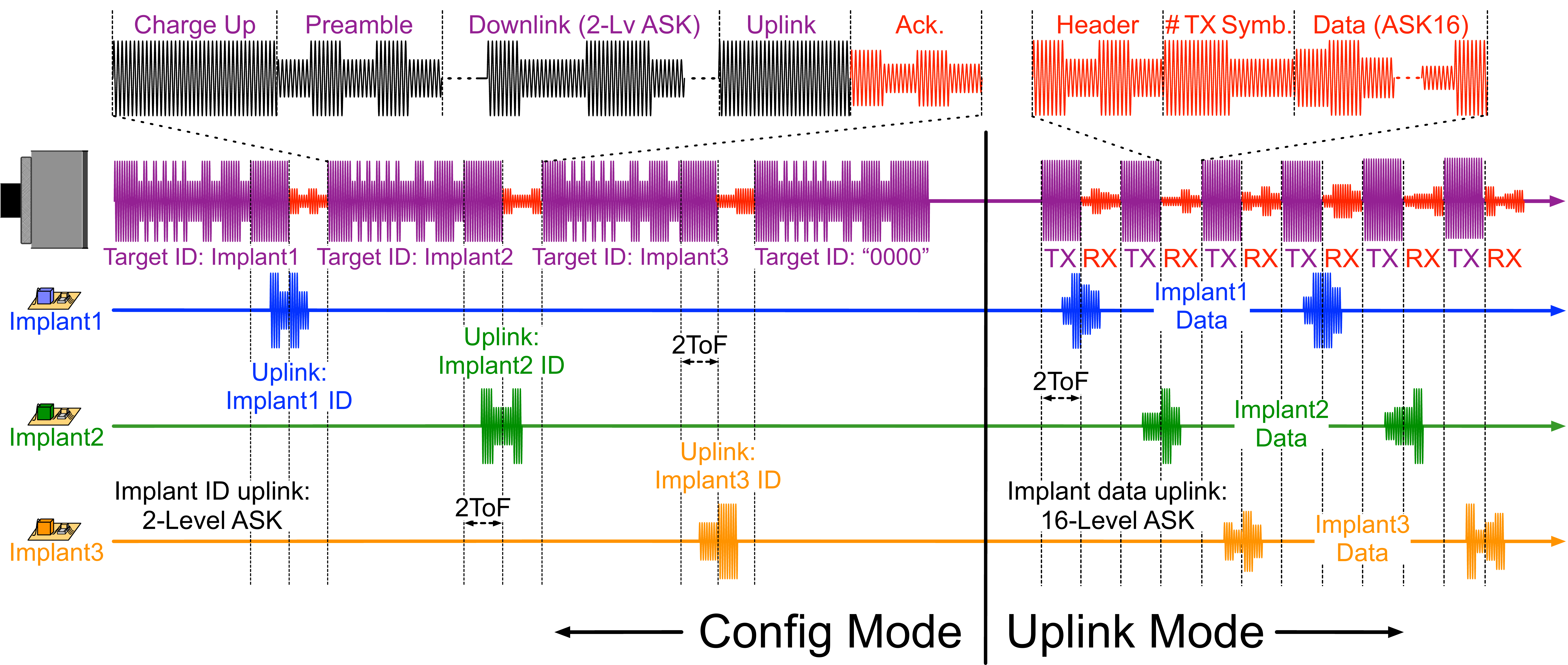}
\setlength\abovecaptionskip{1pt}
\caption{Complete timing diagram of the DustNet communication protocol. In Config Mode, the external transducer transmits pulses that encode link parameters and a target implant ID. Implants matching the target ID backscatter the ID to acknowledge successful configuration. A transmitted target ID of 0 signals implants to transition to Uplink Mode, which implements the TDMA protocol for data transmission.}
\vspace{-8pt} 
\label{fig_timing_diagram}
\end{figure*}

Multi-level ASK modulation can be easily implemented using a resistive digital-to-analog converter (R-DAC) as the termination impedance. However, linearly spaced R-DAC resistances would result in nonlinear spacing of the received backscatter amplitude; this compresses the ASK constellation and reduces $d_{min}$, thus increasing the BER due to carrier and electronic noise, as shown in Fig. \ref{fig_mASK}(a). Furthermore, if nonlinearly spaced resistances were used to generate linear echo levels, variations in R-DAC levels would be amplified by the nonlinear relationship between Γ and $Z_E$. To minimize the system BER, linearly spaced echo levels are required. Analytical optimization demonstrates that even an ideal passive network compresses $d_{min}$ by $1.85\times$ compared to linear ASK level spacing across the same restricted dynamic range; this distortion corresponds to an SNR penalty of 5.3dB. To create linearly spaced backscatter levels, we can instead modulate the voltage across the piezo terminals. Rewriting Equations (\ref{firsthalf}) and (\ref{secondhalf}) demonstrates that Γ is linear in the piezo voltage $V_{PZ}$ (Equations (\ref{eq3}-\ref{eq4})). Because the piezo impedance is purely resistive at resonance, Γ is linearly related to the current across the piezo terminals. Therefore, linearly spaced backscatter levels can be easily realized using a current DAC (I-DAC), as shown in Fig. \ref{fig_mASK}(b). The I-DAC can easily be made programmable to account for variations in piezo impedance. Because the effective resistance of the piezo is higher at parallel resonance, the carrier frequency $f_c$ of the transmitted ultrasound is chosen to match $f_p$ of the implant piezo to achieve larger modulation of Γ. If $f_c$ does not match $f_p$, the impedance of the piezo will no longer be purely resistive; the added reactance will cause the piezo voltage - and therefore Γ - to settle more slowly, which may cause inter-symbol interference if the symbol width is too short. The symbol width is therefore configurable to maintain a low BER in the presence of piezo $f_p$ variation. In this scheme, the number of ASK levels is limited by the carrier, I-DAC, and channel noise.


\begin{table}[]
\centering
\caption{Programmable Uplink Parameters}
\label{table01}
\renewcommand{\arraystretch}{1.5}
\setlength\tabcolsep{1.5pt}
{\scriptsize 
\begin{tabular}{|m{0.15\textwidth}<{\centering}|m{0.15\textwidth}<{\centering}|m{0.15\textwidth}<{\centering\arraybackslash}|}
    \hlineB{2}\textbf{}
    \textbf{Parameter} & \textbf{Range} & \textbf{Description} \\
    \hline
    I-DAC Unit Current & 4 - 40µA & Account for piezo resistance variations\\
    \hline
    Number of Uplink Samples per Packet & 1 - 16 & Enables varying implant depths\\
    \hline
    ASK Levels & 2, 4, 8, 16 & Adjust data rate for noisy channels\\
    \hline
    Total Number of Implants & 1 - 8 & Communicate with multiple implants\\
    \hline
    Uplink Index & 1 - 8 & Sets order of channel control in Uplink Mode\\
    \hline
    LFSR Enable & 1 or 0 & Enable LFSR for BER characterization\\
    \hline
    Number of Cycles per Symbol & 4, 6, 8, 10, 12, 14, 16 & Adjust symbol rate for piezo settling time\\
    \hline
    ADC Slice Selection & 0-8; 1-9; 2-10; 3-11 & Sets ADC range \\

\hline




\end{tabular}}

\end{table}
\subsection{Communication Protocol Implementation}

To enable high-speed communication and link configuration, DustNet implements two modes of operation: Config Mode and Uplink Mode. Upon powering up, all implants first enter Config Mode. In Config Mode, the external transducer encodes data in the amplitude of the transmitted wave to program link parameters for each implant. As shown in Fig. \ref{fig_timing_diagram}, Config Mode pulses consist of four sections: (1) charge-up, (2) preamble, (3) downlink data, and (4) uplink. The charge-up section is transmitted at maximum amplitude and is used by each implant to charge its storage capacitor. During the preamble, the amplitude of the incoming US wave is modulated by the external interrogator using 2-level ASK. The pattern “10” is repeated 32 times so that the implants can estimate the downlink symbol width. The number of ultrasound cycles transmitted per downlink symbol is configurable to ensure proper implant configuration in the presence of transducer or piezo variations. Symbol width is estimated by averaging the number of ultrasound cycles transmitted in each “1” and “0”. At the end of the preamble, a “11001100” header pattern is transmitted to indicate the end of the preamble. The subsequent downlink data section consists of 48 Manchester encoded link configuration bits. The implants use the previously measured symbol width to sample each symbol at its midpoint and compare the sampled voltage to the average envelope voltage to extract the digital symbol; Manchester encoding ensures that the average envelope voltage remains constant. To indicate which implant is being configured, the first 8 bits are reserved for a target implant ID. Each implant has a 3-bit hard-coded ID assigned by tying ID pads on the DustNet IC to $V_{DD,D}$ or GND; if the implant ID matches the transmitted target ID, the implant stores the new link parameters. A list of configuration parameters programmed in Config Mode is enumerated in Table I. Once all the configuration bits are sent, the last portion of the Config Mode pulse is reserved for implant uplink. If the target ID matches the hard-coded implant ID, the implant will backscatter a “0101” header followed by its own ID to acknowledge successful configuration. This also enables the external transducer to discover how many implants are in the system by sweeping all available implant IDs and observing which IDs receive a response.

Once all implants have been configured, the DustNet system moves to Uplink Mode. The external transducer indicates the start of Uplink Mode by transmitting a reserved chip ID (“00000000”) in the final Config Mode pulse; all subsequent pulses are Uplink Mode pulses, which consist of a short charge-up segment, a header segment, and a data segment. To limit requirements on the dynamic range or TX/RX isolation of the external interrogator, the duration of Uplink Mode pulses are set to twice the length of the ultrasound time-of-flight (ToF) between the interrogator and implant, as shown in Fig.~\ref{fig_timing_diagram}. This prevents TX/RX packet overlap, so that the interrogator operates in either TX or RX mode at any given time. Uplink Mode pulse widths can also be adjusted for each pulse in the TDMA protocol to account for different implant depths and can be shortened to accommodate TX ring-up/down. As in Config Mode, implants use the charge-up segment to charge their storage capacitor. During the header segment, implants use the uplink modulator to transmit a “1010” header and the number of data samples transmitted in the Uplink Mode pulse on the amplitude of the backscattered wave using 2-level ASK modulation. The header is followed by data encoded using $2^M$-level ASK modulation, where M is a configurable integer from 1-4 and represents the number of bits transmitted per symbol. The number of US cycles transmitted per symbol ($N_{CpS}$) is programmable in Config Mode. As shown in the Uplink Mode timing diagram in Fig.~\ref{fig_timing_diagram}, only one implant backscatters data during each Uplink Mode pulse; the order in which implants transmit data is programmed in Config Mode. The implants keep track of channel control by counting the number of received Uplink Mode pulses and only transmitting data during their designated slot.

To ensure lossless data transmission, the rate of data generation of each implant cannot exceed its allocated share of the overall transmission bandwidth. Each implant generates data at a rate of $f_sN_{bit}$, where $f_s$ is the sampling frequency and $N_{bit}$ is the number of bits per sample. During an uplink pulse, the maximum transmission data rate is given by $\frac{f_cM}{N_{CpS}}$, where $f_c$ is the US carrier frequency. Due to the pulse-echo communication scheme, the implants are only transmitting with 50\% duty cycle, resulting in a total available data rate of $DR_{total} = \frac{1}{2}\frac{f_cM}{N_{CpS}}$. Assuming the bandwidth is divided equally among $N_{imp}$ implants, the system must satisfy the condition presented in Equation \ref{ineq_DR}:

\begin{equation*}
\refstepcounter{equation}\latexlabel{ineq_DR}
f_sN_{bit} \leq \frac{1}{2}f_c\frac{M}{N_{CpS}}\frac{1}{N_{imp}}
\tag{\ref{ineq_DR}}
\end{equation*}

\section{Circuit Implementation}

\begin{figure}[t]
\centering
\includegraphics[width=8.0Cm]{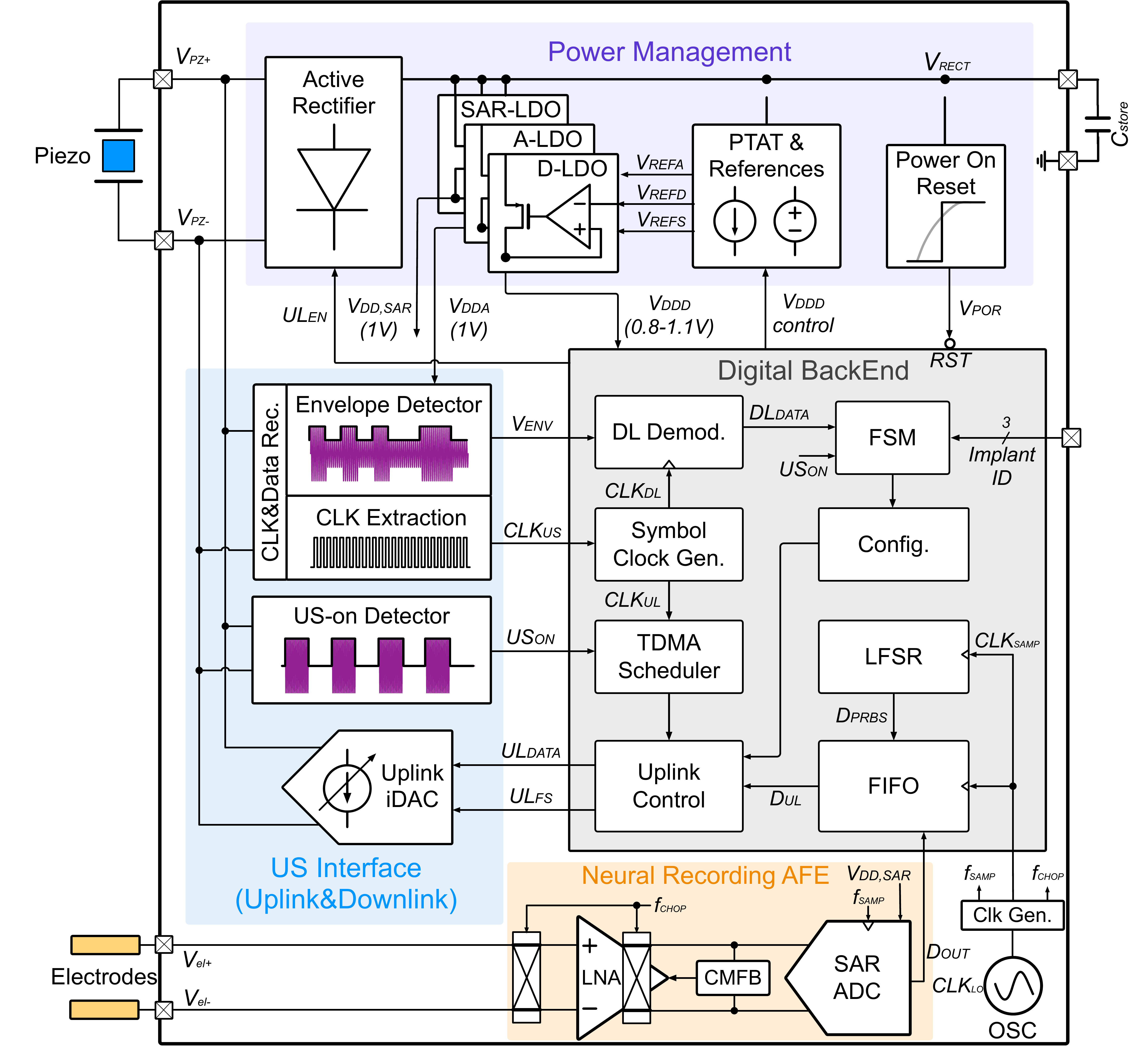}
\setlength\abovecaptionskip{1pt}
\caption{DustNet IC block diagram, including power management blocks, the neural recording AFE, a digital backend, and the ultrasound interface. The electrodes, implant piezo, storage capacitor, and hard-wired ID fuses are externally connected to pads on the IC.}
\label{fig_block_diagram}
\vspace{-12pt}
\end{figure}

Fig. 6 shows the block diagram of the DustNet IC. The IC has 5 main blocks: (1) power management, including an active rectifier and low-dropout regulators (LDOs); (2) an ultrasound interface with downlink data processing and a linear, multi-level ASK uplink modulator; (3) a low-noise neural recording analog front-end (AFE); (4) on-chip clock generators; and (5) a digital backend to manage the communication protocol. During operation, the implants continuously record neural data using the AFE and store it on an on-chip first-in first-out (FIFO) memory. Once powered on by the first incoming US pulse, the implants do not turn off; pulses are structured such that the implants can store enough energy to stay powered during inter-pulse gaps. 

\subsection{Power Management and Integrated Uplink Modulator}

\begin{figure}[t]
\centering
\includegraphics[width=8.5Cm]{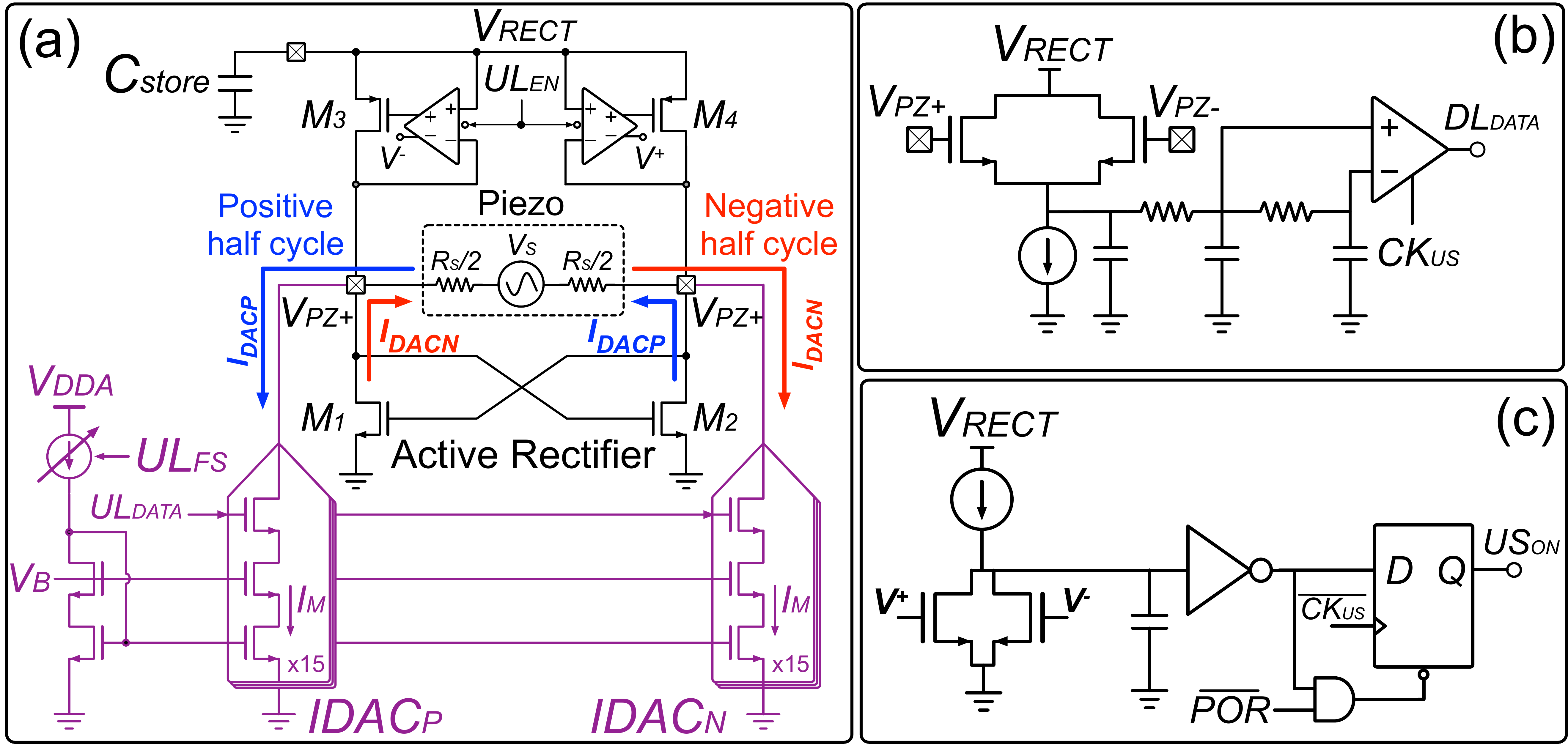}
\setlength\abovecaptionskip{1pt}
\caption{(a) Active rectifier with integrated linear 16-level uplink modulator. (b) Envelope detector circuitry to process downlink data. (c) Ultrasound-on detector circuitry to detect the start of TX pulses.}
\label{fig_schematics}
\end{figure}

The active rectifier converts the AC voltage transduced by the piezo into a DC voltage ($V_{rect}$) that is stored on a 100 nF capacitor ($C_{store}$). $V_{rect}$ varies with the amplitude of the received US wave and drops during the gap between US pulses, but always exceeds 1.2V after start-up to ensure stable system functionality; the size of $C_{store}$ is chosen such that $V_{rect}$ remains above 1.4V when the ultrasound is off during the pulse-echo communication protocol. LDOs regulate the variable $V_{rect}$ to create stable voltages for the analog ($V_{DD,A}$, 1V) and digital ($V_{DD,D}$, 0.8V) supply domains, as well as a separate 1V supply for the SAR ADC. The LDOs implement common-gate compensation to ensure stability during low load conditions at implant start-up. To ensure clean supplies during fast transient load steps (e.g., digital switching, SAR ADC sampling), the LDOs rely on integrated on-chip load capacitance ($C_L$). To maximize transient load regulation, the available unused silicon area was filled with decoupling capacitance. A supply-independent, proportional-to-absolute-temperature (PTAT) reference circuit provides voltage references ($V_{REF,A}, V_{REF,D}$) for the LDO blocks. Once the LDO outputs are settled, a power on reset (POR) block raises a flag ($V_{POR}$) to activate the digital backend and begin data sampling.

The uplink modulator consists of two 16-level linear I-DACs ($I_{DACP}$ and $I_{DACN}$) integrated into the active rectifier topology, as shown in Fig. \ref{fig_schematics}(a). When the rectifier is actively harvesting power from the piezo, the I-DACs are disabled and pass transistors $M_3$ and $M_4$ are toggled such that the positively charged piezo terminal is connected to $C_{store}$ if $|V_{PZ}| > V_{RECT}$. During data uplink, the digital backend turns off $M_3$ and $M_4$ to protect the stored charge on  $C_{store}$ from the large I-DAC currents. The I-DACs consist of 15 identical, switchable current sources that can each pull a current $I_M$ through the piezo. The number of enabled sources corresponds to the amplitude of the backscattered symbol. Using an adjustable current mirror sourced from a PTAT reference, $I_M$ is configurable from 4µA to 40µA to account for variations in the piezo impedance across implants. $I_{DACP}$ conducts current during the positive half-cycle of the incoming US wave, and $I_{DACN}$ conducts current during the negative half-cycle to preserve the sign of Γ modulation. The current mirror ratio is large to ensure low quiescent power consumption. Since the chip is not actively sourcing current into the piezo, the current sources are equivalent to a variable termination impedance that is modulated to ensure that a fixed current flows through the piezo terminals and generate linear echo levels; the modulation current flows through the piezo terminals and is driven by the incident acoustic wave rather than the stored charge on $C_{store}$.

The active rectifier topology used in this work is based on that presented in \cite{Ghanbari_JSSC19}, which reports a measured efficiency of 80\% under similar operating conditions; although the rectifier in this work integrates the uplink modulator, the uplink modulator is disconnected during rectification and does not affect efficiency. Overall power harvesting efficiency must also consider the nature of the pulse-echo communication scheme: the implant cannot harvest power during the interrogator RX phase of each pulse-echo cycle, which alone limits power harvesting to 50\% of the total channel time. Furthermore, the implant cannot harvest power while actively modulating data during its assigned TDMA uplink slot. The overall power harvesting duty cycle can then be calculated using the per-implant data rate and communication parameters: with a per-implant data rate of 50 kb/s, a symbol duration of 8 carrier cycles, 16-level ASK modulation transmitting 4 bits per symbol, and a 2 MHz carrier frequency, each implant spends 50 ms out of every second transmitting data, leaving 450 ms (45\% of total channel time) available for harvesting. The combined effect of rectifier efficiency and power harvesting duty cycle yields an overall efficiency of 36\%, implying that the piezo must transduce an average of 19.4 µW of electrical power to sustain the 7 µW implant power - a requirement that is well within FDA acoustic safety limits at 90mm depth.

\subsection{Ultrasound Interface \& Digital Backend}
The communication protocol and data sampling are managed by the digital backend. After $V_{POR}$ is raised during start-up, the digital backend begins to write data from the ADC ($D_{OUT}$) or LFSR ($D_{PRBS}$) into the 16 x 9-bit FIFO memory. To ensure that no data is overwritten, the FIFO depth $D_{FIFO}$ must satisfy: $D_{FIFO} \geq N_{imp}T_{pulse}f_s$, where $T_{pulse}$ is the duration of the Uplink Mode pulses. The 6.25 kHz $CLK_{samp}$ serves as the write enable for the FIFO memory and advances the LFSR code. 

In Config Mode, an envelope detector circuit processes the symbols transmitted by the external interrogator ($V_{ENV}$), which are demodulated and processed by a finite-state machine (FSM). To identify the downlink symbols ($DL_{DATA}$), the envelope detector compares the instantaneous US envelope voltage with the average envelope voltage using two low-pass filters with different corner frequencies, as shown in Fig. \ref{fig_schematics}(b). The FSM processes the downlink data and updates the link parameters stored in configuration registers if the transmitted target ID matches the externally hard-coded implant ID set using pads on the IC. 

During Uplink Mode, 8 or 9 bits of the FIFO output ($D_{UL}$) are selected and assembled into M-bit transmission packets ($UL_{DATA}$) by the uplink control module for $2^M$-level ASK communication; 8 bits are selected for 2,4, and 16-level ASK modulation, while 9 bits are selected for 8-level ASK modulation to ensure an integer number of symbols per data sample. The bits selected from the 12-bit ADC output are customizable in Config Mode; this enables a configurable ADC range to account for varying signal amplitudes and noise floors while maintaining an achievable per-implant data rate. $UL_{DATA}$ is used as the I-DAC code in the uplink modulator.  The implants count the Uplink Mode pulses using the US-On block to monitor which implant is in control of the channel. A schematic of the US-On circuitry is shown in Fig. 7(c); $V^+$ and $V^-$ are the inverting outputs of the rectifier comparators in Fig. \ref{fig_schematics}(a). When the implant is receiving US power, the NMOS transistors are active and discharge the capacitor, raising the $US_{ON}$ flag; when the US is off, $V^+$ and $V^-$ are pulled down and the capacitor is recharged by the current source. The FSM counts the $US_{ON}$ signals and uses the uplink index parameter to determine if the implant should transmit data in the current Uplink Mode pulse. If the implant controls the channel, the uplink modulator is activated to perform multi-level ASK backscatter. To evaluate the BER of the communication protocol, a 16-bit LFSR is implemented in the digital backend. If activated during Config Mode, the LFSR block generates a pseudorandom binary sequence (PRBS) that is written to the FIFO memory. To calculate the system BER, the ASK-modulated data sequence received over the US link is compared to the PRBS reference.

\subsection{Neural Recording AFE}

\begin{figure}[t]
\centering
\includegraphics[width=8.0Cm]{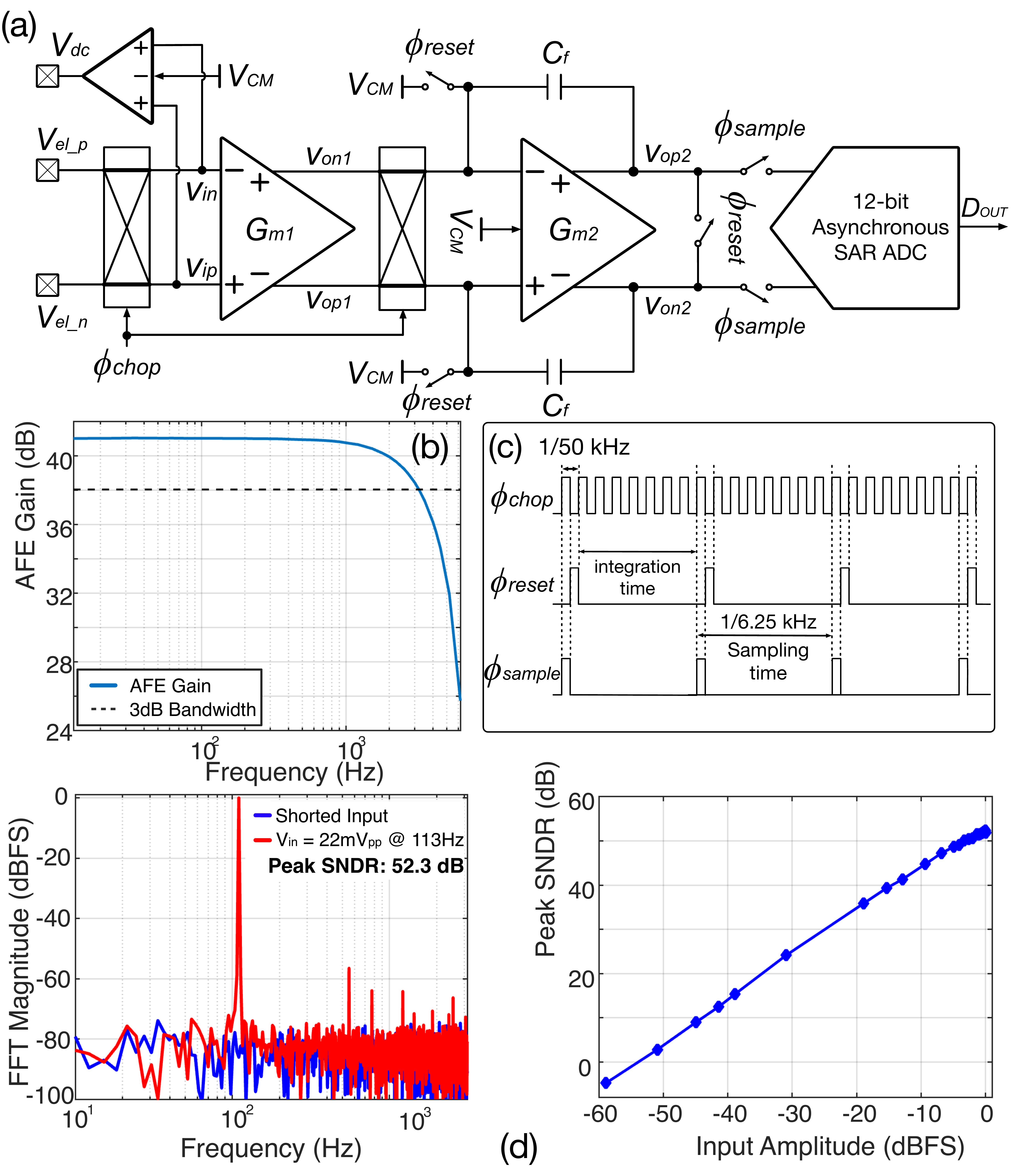}
\setlength\abovecaptionskip{1pt}
\caption{(a) AFE signal chain consisting of a chopped low-noise $G_m$-cell, switched-capacitor integrator, and 12-bit asynchronous SAR ADC. (b) Measured AFE output spectrum. (c) AFE timing diagram. (d) Measured AFE spectrum at maximum SNDR with $V_{in}$ = 22$mV_{pp}$ (left, red), noise spectrum (left, blue) and SNDR as a function of input amplitude (right).}
\label{fig_ADC}
\vspace{-15pt}
\end{figure}

To record neural signals (e.g., LFPs, electromyography (EMG) signals, etc.), the DustNet IC implements an analog front-end that consists of a low-noise amplifier and a 12-bit asynchronous SAR ADC. A schematic-level diagram of the neural recording AFE is shown in Fig. \ref{fig_ADC}(a). The low-noise amplifier consists of a chopped $G_m$ stage ($f_{chop}$ = 50 kHz) followed by a switched-capacitor (SC) integrator. The integrator output is sampled by the SAR ADC at 6.25 kS/s to enable the $>3kHz$ bandwidth required for peripheral nerve recording applications. The open-loop AFE architecture is designed to maximize input impedance and minimize power consumption compared to closed-loop topologies; although it is susceptible to PVT-induced gain variation, absolute gain precision is not required for many neural decoding applications. Depending on the number of ASK levels, a configurable set of 8 or 9 bits from the ADC output are stored in the FIFO buffer to be transmitted back to the external interrogator, yielding a per-implant data rate of 50 or 56.25 kb/s, respectively.

The first $G_m$ stage consists of a fully-differential OTA, which is used for its high speed and noise efficiency. A common-mode feedback (CMFB) circuit sets the output DC voltage to match a reference supplied by an on-chip LDO. The current from the $G_m$ cell is integrated on a feedback capacitor $C_f$ in the second stage. The SC integrator consists of a two-stage, Miller-compensated OTA with CMFB. The AFE bandwidth is set by the integration time of the SC amplifier; because one cycle is used for reset, the integration time ($T_s$) is 7 $CLK_{LO}$ cycles (140µs). The sinc transfer function of the integrator inherently has a 3-dB bandwidth of $\frac{e}{2\pi T_s} = 3.16kHz$, eliminating the need for explicit anti-aliasing filters for high-frequency noise. The measured AFE gain as a function of frequency is shown in Fig. \ref{fig_ADC}(b). The overall signal DC gain ($H_0$) is designed to be 100 to suppress the ADC quantization noise. Previously fabricated US-based implants using ENIG-based electrodes have demonstrated DC offsets of up to 1.5mV \cite{Ghanbari_JSSC19}. Therefore, the full-scale range of the ADC is designed to be 2V with a large differential input range of approximately 20mV to ensure that neural signals can still be recorded in the presence of large electrode DC offsets.

A clock generation circuit provides a stable 50 kHz reference ($CLK_{LO}$) using an offset-compensated RC relaxation oscillator \cite{Paidimarri16}. An on-chip oscillator is used instead of dividing the extracted US clock to allow the carrier frequency to be tuned to the piezo resonant frequency without altering the data sampling rate. $CLK_{LO}$ is directly used as $\phi_{chop}$ to chop the AFE inputs, and is divided by 8 to generate the sampling clock ($\phi_{sample}$) used for ADC or linear feedback shift register (LFSR) sampling. The ADC samples the output of the integrator at 6.25 kHz ($f_s = f_{chop}/8$). A timing diagram of the chop, reset, and sample clocks is shown in Fig. \ref{fig_ADC}(c); the integrator resets one clock cycle ($\phi_{reset}$) after the ADC samples its output. Because $CLK_{LO}$ controls the sampling frequency of the ADC and SC integrator, variations in the oscillator frequency directly correspond to variations in the AFE gain and implant transmit data rate. Measurement of 9 chips demonstrates a mean oscillator frequency of 51.06kHz with a standard deviation of 1.5\% without calibration. Because the AFE gain is inversely proportional to the oscillator frequency, this approximately corresponds to a highly tolerable gain variation of 1.5\% across implants. Furthermore, the corresponding 1.5\% variation in transmit data rate is easily handled by the TDMA protocol if Equation 5 is satisfied.

AFE measurement results are shown in Fig. \ref{fig_ADC}(d) using the 9-bit ADC output from the chip. The differential input range of the ADC is measured to be 22$mV_{pp}$. With a 113Hz full-scale input tone, the AFE achieves a peak signal-to-noise and distortion ratio (SNDR) of 52.3 dB, yielding an effective number of bits (ENOB) of 8.4 bits. The dynamic range is measured to be 53.8 dB, which is suitable for neural recording applications. The effective range of the ADC can also be adjusted by configuring the transmitted bits in Config Mode (e.g., ADC MSBs or ADC LSBs). By transmitting the 9 least significant bits of the ADC output, the input-referred integrated noise of the combined amplifier and ADC is measured to be 9.8 µ$V_{rms}$ across the 3.125 kHz bandwidth.

\vspace{-8pt}

\begin{figure}[t]
\centering
\includegraphics[width=8.0Cm]{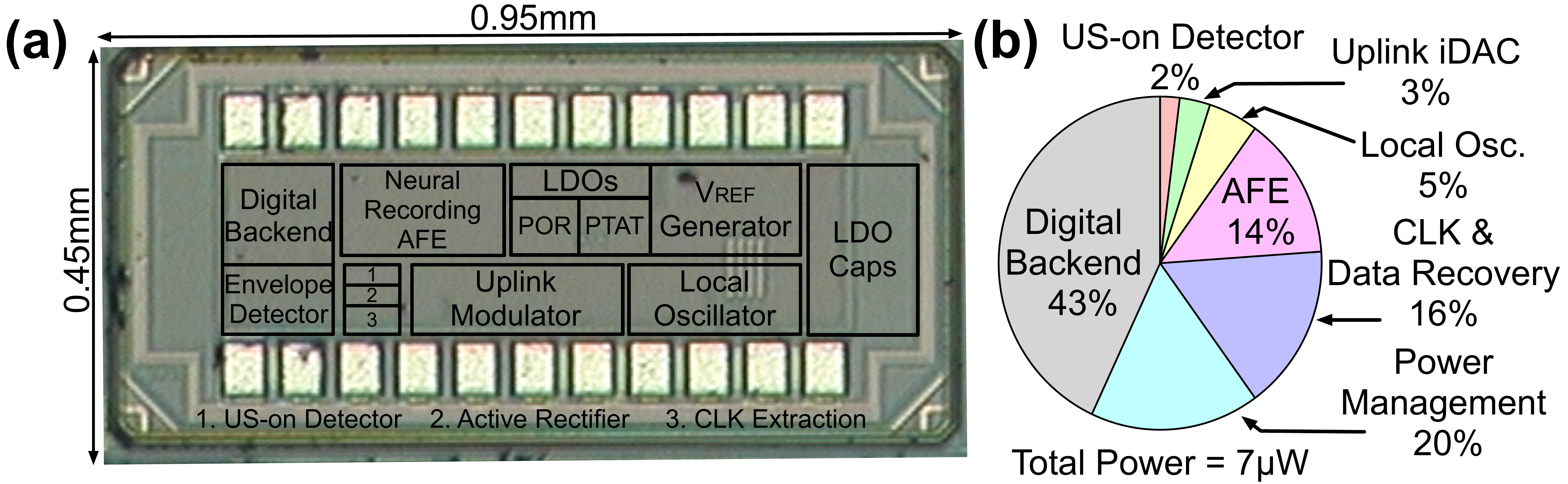}
\setlength\abovecaptionskip{1pt}
\caption{(a) DustNet IC micrograph highlighting the location of various blocks. The chip occupies a footprint of 0.43 $mm^2$. (b) Power breakdown of the IC. The implant consumes a total power of 7 µW.}
\label{fig_micrograph}
\end{figure}

\begin{figure}[t]
\centering
\includegraphics[width=8.5Cm]{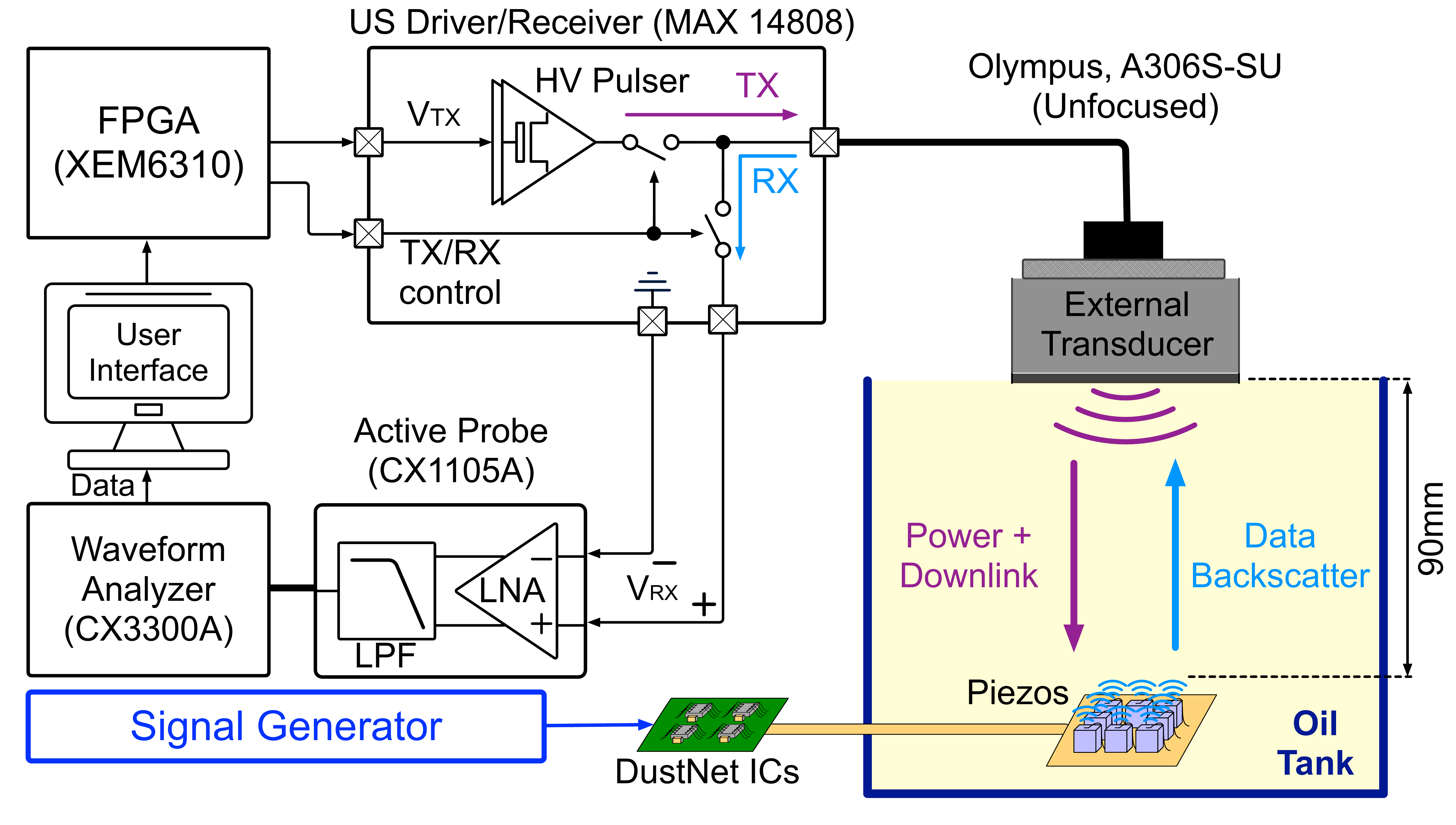}
\setlength\abovecaptionskip{1pt}
\caption{\textit{In vitro} measurement setup. Implant piezos are mounted on an fPCB submerged 90mm in canola oil and connected via external wires to the DustNet ICs. An FPGA and high voltage US pulser board are used to generate control signals for the external interrogator, and a signal generator supplies pre-recorded neural signals to the ICs.}
\label{fig_measurement_setup}
\end{figure}

\begin{figure}[t]
\centering
\includegraphics[width=8.0Cm]{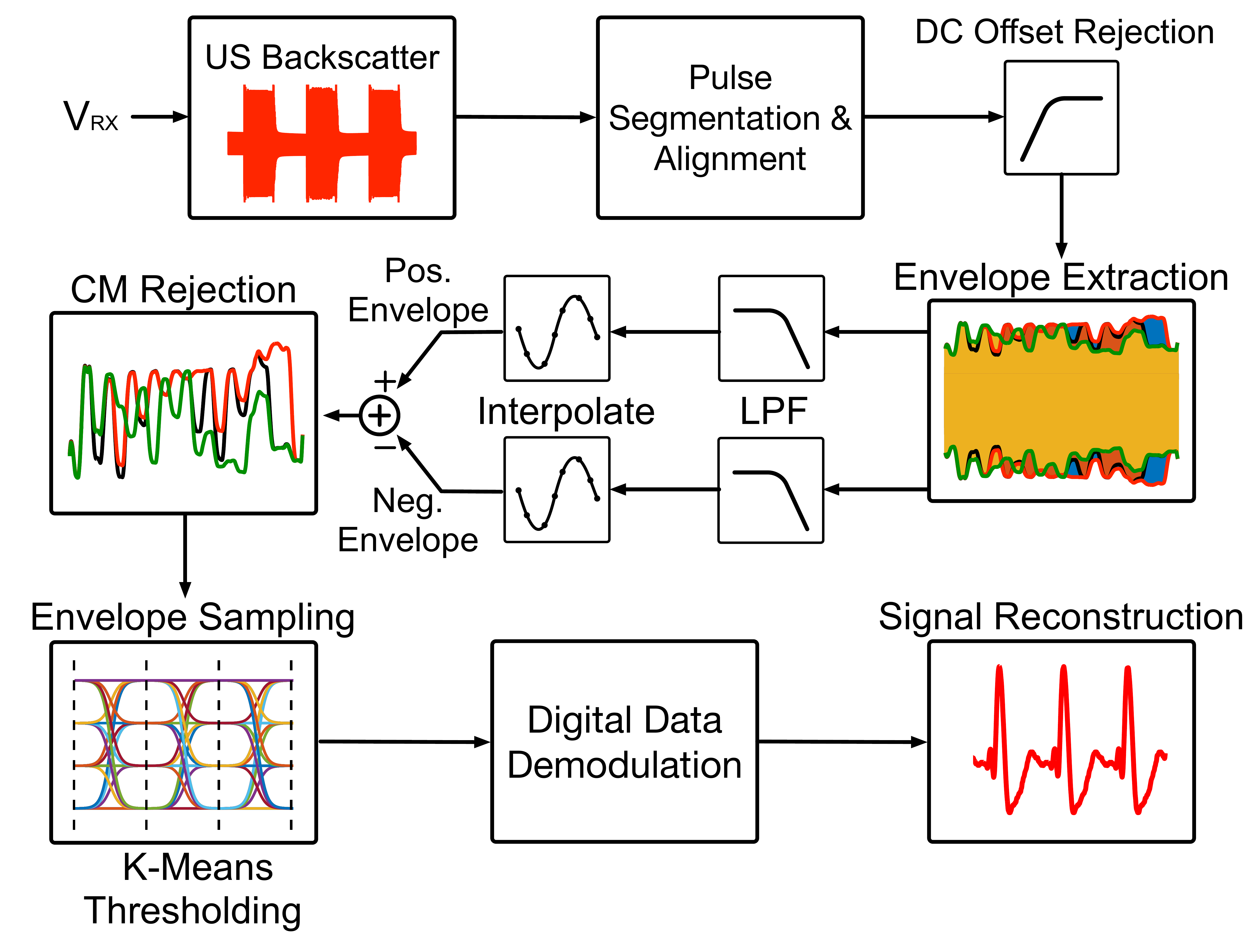}
\setlength\abovecaptionskip{1pt}
\caption{Signal processing chain to demodulate data encoded in the US backscatter. $V_{RX}$ is recorded directly from the output of the HV pulser board. After demodulation of the individual pulses, data are concatenated to reconstruct the recorded ADC/LFSR output.}
\label{fig_sp_chain}
\end{figure}

\begin{figure}[!ht]
\centering
\includegraphics[width=8.0Cm]{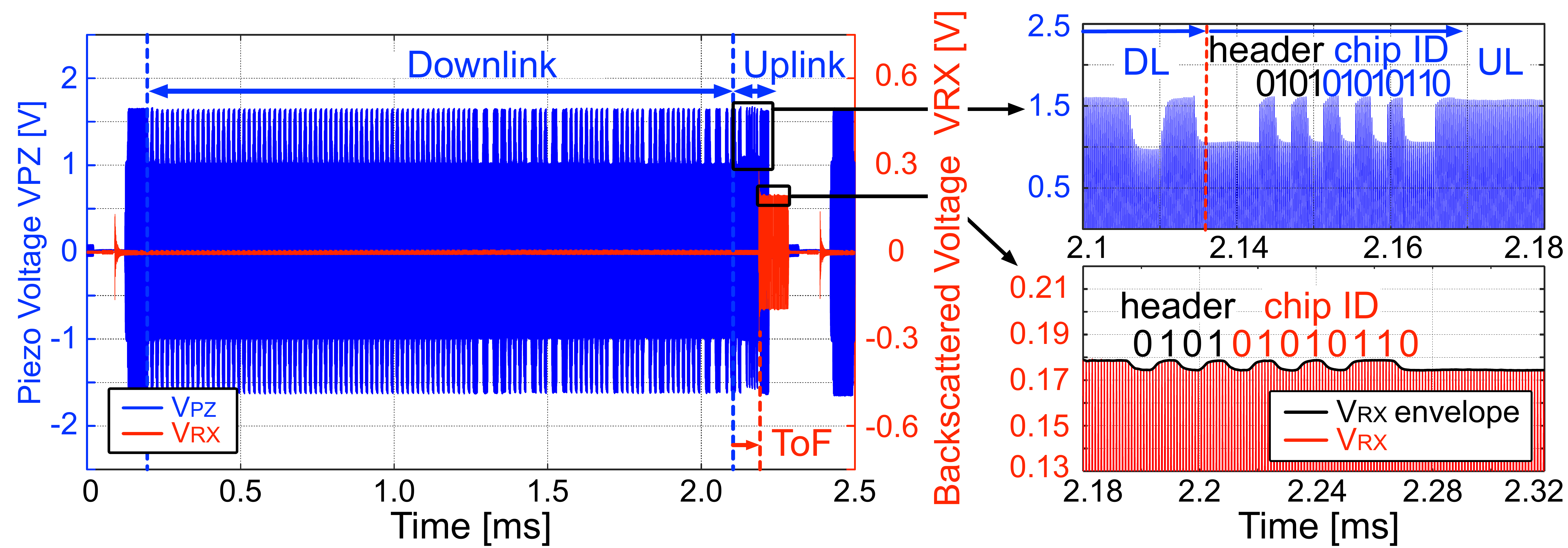}
\setlength\abovecaptionskip{1pt}
\caption{Measured Config Mode waveform at the implant piezo (blue) and implant backscatter received at the external interrogator (red). Because the target ID matches the implant ID, the implant backscatters its own ID, which arrives at the external interrogator after 1 ToF.}
\label{fig_config_mode_wv}
\end{figure}

\section{Measurement Results}

The DustNet IC was fabricated in a 28nm CMOS process. A chip micrograph and power breakdown is shown in Fig. \ref{fig_micrograph}. The functionality of the DustNet system was verified at 90mm depth in canola oil, which has an acoustic attenuation coefficient ($\sim$0.25 dB/cm/MHz \cite{rabbani2021}) similar to human tissue \cite{abbott1999}.

\begin{figure*}[t]
\centering
\includegraphics[width=18Cm]{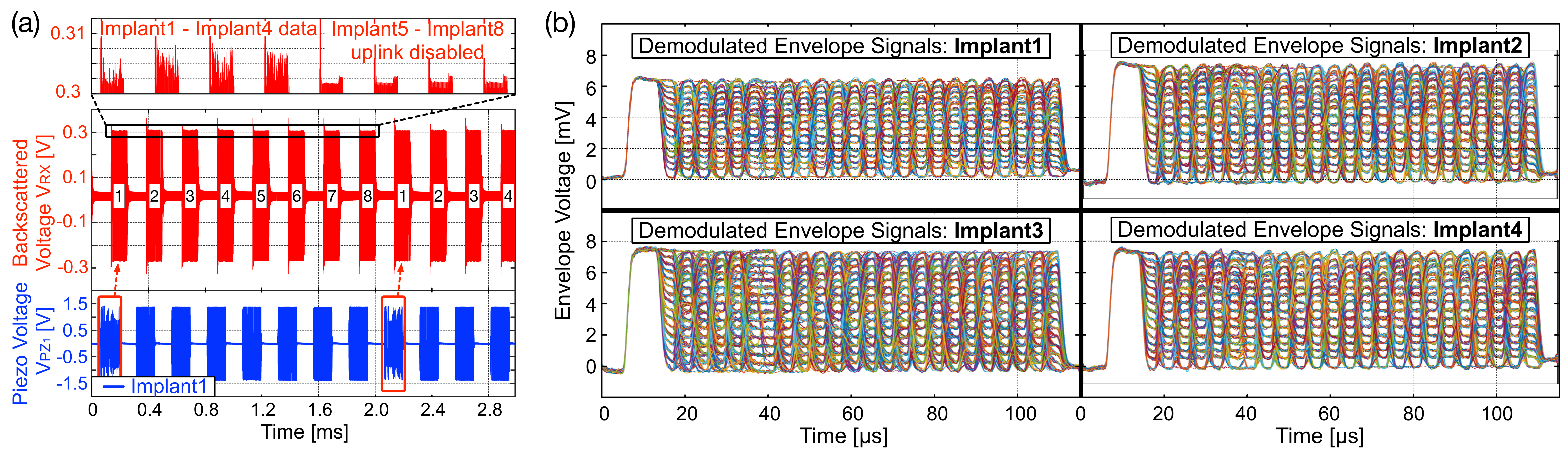}
\setlength\abovecaptionskip{1pt}
\caption{(a) Backscattered Uplink Mode waveforms received at the external interrogator (red) when the system is configured to support 8 implants with 4 implants enabled, and the piezo voltage of implant 1 (blue) demonstrating the TDMA protocol. The backscatter amplitude is small compared to the large background reflection from the fPCB. (b) Demodulated eye diagrams for four implants transmitting data using 16-level ASK modulation over the ultrasound link. The received echo levels are clearly separable, indicating that no errors were made.}
\label{fig_multi_implant_test}
\vspace{-12pt}
\end{figure*}

\begin{figure}[t]
\centering
\includegraphics[width=8.0Cm]{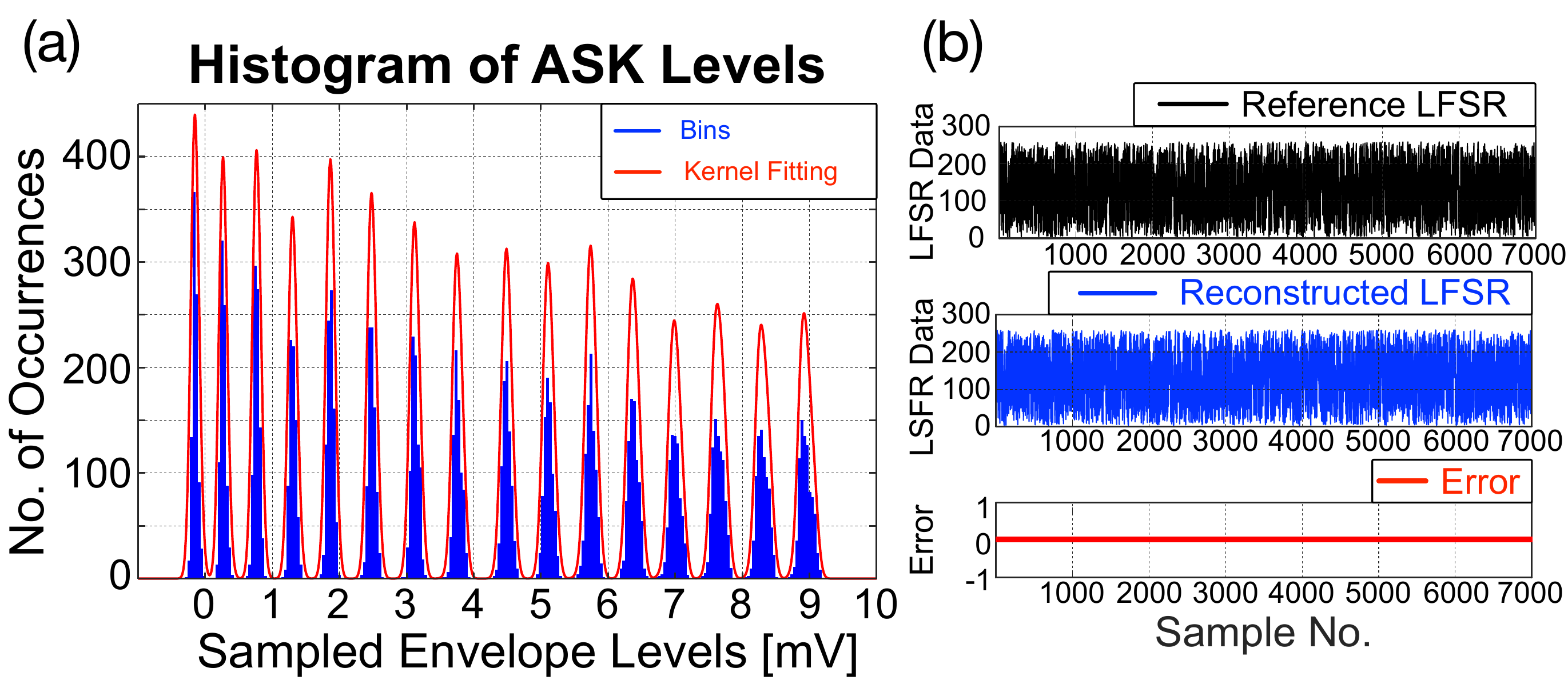}
\setlength\abovecaptionskip{1pt}
\caption{(a) A histogram of envelope voltages extracted from 827 Uplink Mode packets from a single implant modulated with LFSR data with a fitted PDF demonstrating 16 separable levels. (b) Reconstructed LFSR data (blue) compared to a golden reference (black). No errors (red) are made when transmitting 52928 bits, yielding a BER of $\leq$1.89E-5.}
\label{fig_single_measurement}
\end{figure}

\subsection{Measurement Setup \& Data Processing}

The measurement setup is shown in Fig. \ref{fig_measurement_setup}. Four miniaturized piezoceramic transducers of size 0.7 x 0.7 x 0.7 mm\textsuperscript{3} with $f_p \approx$ 2~MHz are mounted on a flexible PCB (fPCB) using conductive silver epoxy in a tank filled with canola oil; a cubic factor is optimal for backscatter communication \cite{Ghanbari_TBIOCAS20}. An acoustic absorber (Aptflex F28P, Precision Acoustics) is placed at the bottom of the tank under the fPCB. An unfocused ultrasound transducer (A306S-SU, Evident Scientific) is vertically mounted in the tank 90mm directly above the transducer. Wires connect four DustNet ICs mounted on external daughterboards to the fPCB in the tank.

A 5-level high voltage pulser (MAX14808, Analog Devices) with a built-in TX/RX switch drives the unfocused US transducer at 2 MHz. An FPGA (XEM6310, OpalKelly) generates control signals for the pulser to form the Config Mode and Uplink Mode packets. To record the US backscatter, the pulser switches to RX mode one ToF after the chip begins to transmit data. The received voltage ($V_{RX}$) is recorded by a low-noise active probe (CX1105A, Keysight) with a 10 MHz low-pass filter (LPF) connected to a high-resolution current waveform analyzer (CX3300A, Keysight).

To decode ASK-modulated data from the received backscatter waveforms, several signal processing techniques are used to extract the US envelope and remove out-of-band noise and other undesired fluctuations. A complete block diagram of all the signal processing steps used to demodulate the US backscatter is shown in Fig. \ref{fig_sp_chain}. First, the timeseries Uplink Mode waveform obtained from the current waveform analyzer is segmented into individual pulses that are time-aligned based on the start of the US pulse. The segments are high-pass filtered using a Butterworth filter with a high pass corner at 20 kHz to remove DC offsets and low-frequency noise. The US envelope is extracted from the peaks and valleys of the recorded US backscatter using spline interpolation. The envelope is then low-pass filtered using a second-order Type 2 Chebyshev filter with a low-pass corner at 10 MHz to eliminate high-frequency noise. By taking the difference between positive and negative envelope waveforms, the system eliminates common-mode noise, which includes multipath effects, passive reflections from implants that are not modulating during the TDMA slot, and slow motion artifacts. This also ensures that symbol decoding is immune to mismatch between the positive and negative I-DAC currents; using both envelopes ensures that the symbol amplitude is proportional to the sum of the I-DAC currents.

The envelope waveforms are sampled at the end of each transmitted symbol and thresholds for each amplitude level are generated using k-means clustering of the received echo voltages. By generating unique thresholds for each amplitude level rather than using a uniform linear spacing to differentiate symbol values, the decoding method is robust to nonlinearities in the I-DACs. The process of generating the thresholds is performed for each implant in the network; this enables the symbol decoding to overcome differences in packet amplitude due to variations in piezo resonance frequency, channel characteristics, or global I-DAC variation across implants. To account for small changes in alignment over time, the 2-level packet header can be used to estimate changes in the full-scale modulation amplitude and recalibrate the thresholds using the relative separation between the 16 levels. To demodulate the data, these thresholds are used to convert each symbol into a 4-bit binary representation, and pairs of transmitted symbols are combined to recover the 8-bit data samples. The system BER is characterized by transmitting PRBS data generated by the on-chip LFSR and comparing the received bitstream to a golden reference.

\subsection{Communication Protocol Verification \& Implant Measurements}

When the US pulser is initially turned on, the DustNet ICs begin to charge their $C_{store}$. $V_{RECT}$ settles to a value $>$1.2V; the achieved voltage varies depending on the US power transmitted by the external interrogator. The regulated supplies $V_{DD,A}$ and $V_{DD,D}$ settle to 1.0V and 0.8V, respectively. Once the LDO voltages have settled, the $V_{POR}$ flag is raised to activate the digital backend.

An example Config Mode waveform ($V_{PZ}$) and the corresponding backscatter received at the external interrogator ($V_{RX}$) are shown in Fig. \ref{fig_config_mode_wv}. $V_{PZ}$ is recorded at the piezo terminals, and the modulated uplink section is received as $V_{RX}$ after 1 ToF. The header and chip ID backscattered by the uplink modulator are clearly seen in both $V_{PZ}$ and $V_{RX}$. Because the carrier frequency is equal to $f_p$, a high $V_{PZ}$ results in a low $\Gamma_p$ and thus a low $V_{RX}$.

To validate the multi-implant TDMA protocol, each implant transmits PRBS data from its on-chip LFSR. A sample set of backscattered Uplink Mode waveforms are shown in Fig. \ref{fig_multi_implant_test}(a), in which the system is configured to support 8 total implants with data transmission enabled for 4 implants. Amplitude modulation for each transmitting implant is clearly demonstrated in the recorded backscatter. The recorded $V_{PZ}$ for implant 1 is also shown in Fig. \ref{fig_multi_implant_test}(a). Because $N_{implants}=8$ is set in Config Mode, each implant backscatters its data once out of every 8 pulses. Eye diagrams for the 16-level ASK backscatter received from each implant using are shown in Fig. \ref{fig_multi_implant_test}(b). The ASK levels are clearly separable, indicating that no errors are made during transmission and that the system would be able to tolerate decreases in packet amplitude due to misalignment. The peak envelope voltage received from each implant varies slightly due to mismatch across the implant ICs and piezos. However, unlike for CDMA protocols, this does not pose an issue to decoding because packets received from different piezos are processed separately, and a set of ASK thresholds are generated for each chip. Each implant transmits 24 4-bit symbols of neural data (96 total bits) during each set of 8 US pulses (duration: 1.9ms). This corresponds to a per-implant data rate of 50kb/s; with 4 transmitting implants, this demonstrates a measured data rate of 200kb/s. Although the experimental setup utilizes fewer physical implants to emulate the 8-slot TDMA frame, this rigorously validates the system's scalability because the TDMA protocol isolates each node's uplink in time and the interrogator processes each timeslot independently. Furthermore, the on-chip FIFO ensures that data generated during each of the idle slots is buffered without data loss, demonstrating that the architecture can support a fully populated 8-node network with 400 kb/s total link data rate. 

A histogram of received backscatter voltages from 827 Uplink Mode packets transmitted by a single implant and a fitted probability distribution (PDF) is shown in Fig. \ref{fig_single_measurement}(a). The voltage levels are clearly separable and the PDF shows 16 distinct peaks, corresponding to the echo amplitudes generated by each of the 16 I-DAC levels. A plot of the reconstructed data and golden reference data is shown in Fig. \ref{fig_single_measurement}(b); there are no errors in the transmitted data stream. The data stream consists of 6616 8-bit samples, yielding a BER of $\leq$1.89E-5.

\begin{figure}[t]
\centering
\includegraphics[width=8.5Cm]{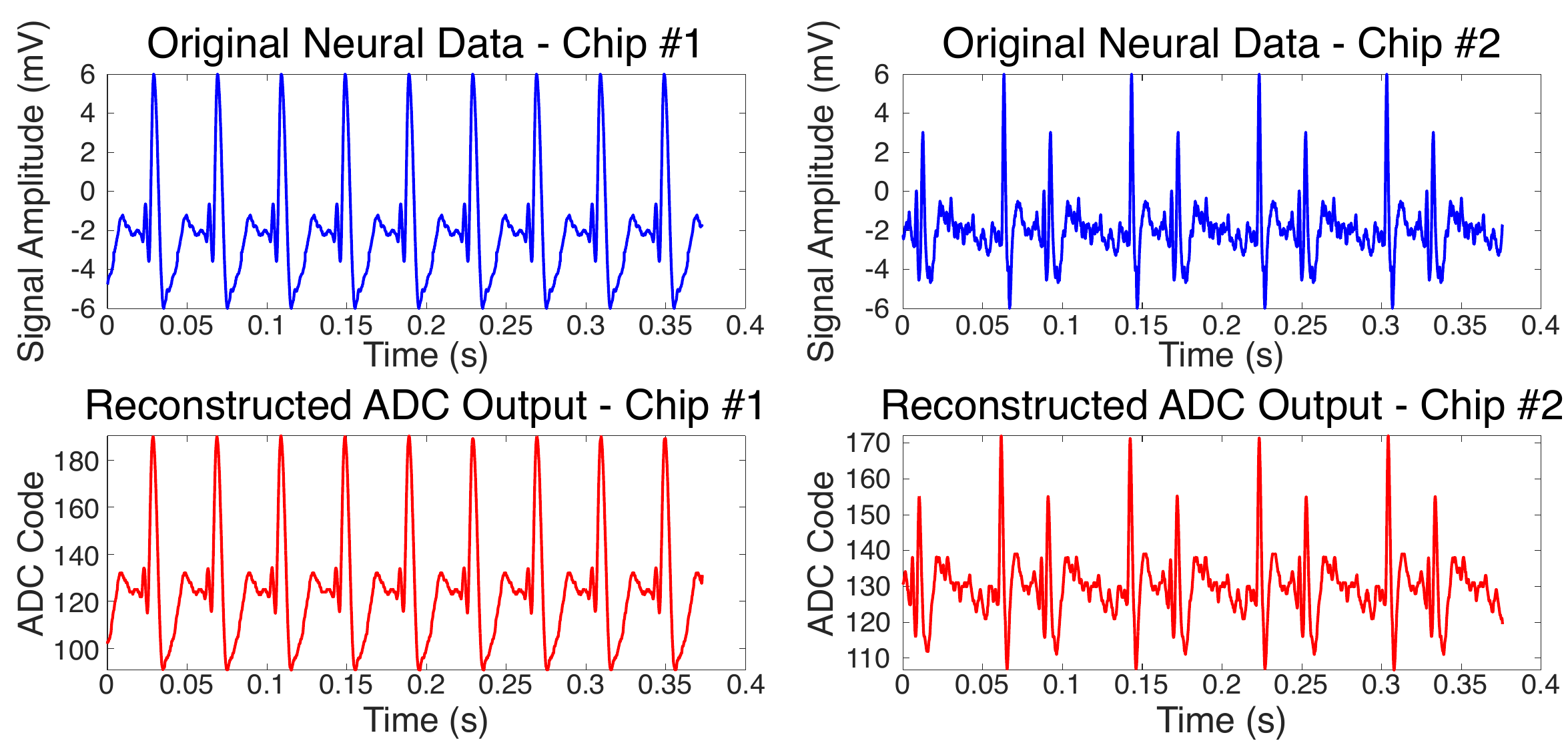}
\setlength\abovecaptionskip{1pt}
\caption{End-to-end system verification. Simulated neural signals (blue) are fed to the implant using a function generator and off-chip electrode model. The signals are quantized by the implant AFE and digital data is transmitted over the US link. Demodulated data (red) closely matches the original neural data.}
\label{fig_ADC_reconstruct}
\end{figure}

To demonstrate end-to-end functionality, the AFE inputs of two implants are connected to an arbitrary function generator programmed to output simulated, slowed neural waveforms. The 8 MSBs of the on-chip ADC outputs are then transmitted over the US link. Fig. \ref{fig_ADC_reconstruct} shows the time-aligned simulated neural waveforms and reconstructed ADC outputs for both implants; the ADC sampling rate is approximately 6.25 kHz for both implants, yielding a data rate of 50 kb/s for each implant. 

\begin{figure}[t]
\centering
\includegraphics[width=8.5Cm]{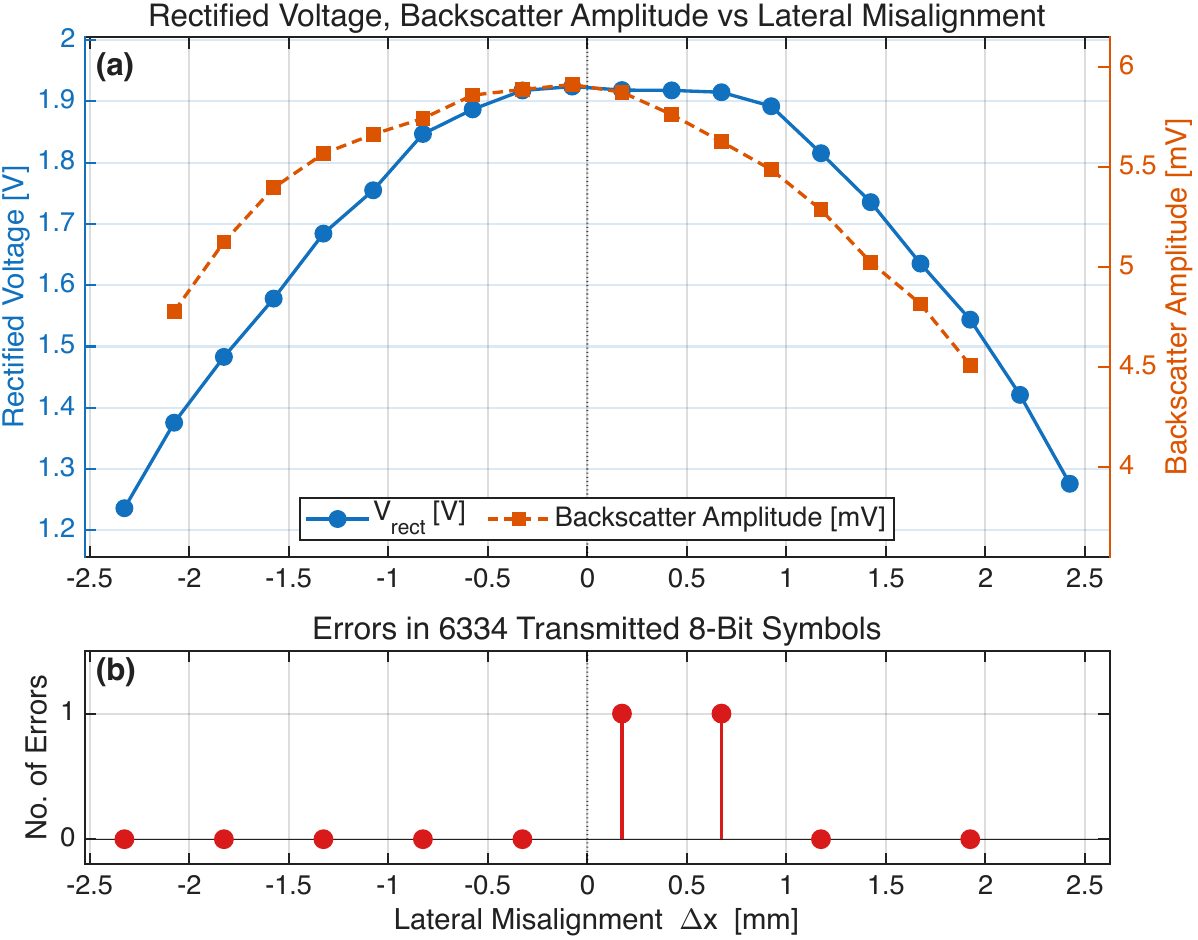}
\setlength\abovecaptionskip{1pt}
\caption{Measured sensitivity to lateral misalignment of a single implant at 90 mm depth in oil. (a) Measured rectified voltage ($V_{RECT}$) and backscatter modulation amplitude as a function of lateral misalignment between the implant piezo and external interrogator. (b) Errors made in 6334 received 8-bit packets; at most one error is made across $\pm$2 mm lateral misalignment.}
\label{fig_misalign}
\end{figure}

The sensitivity of DustNet to lateral misalignment between the external interrogator and the implant was characterized by measuring $V_{RECT}$ and the maximum backscatter modulation amplitude from a single piezo as a function of lateral offset; the performance of the communication protocol was also evaluated by transmitting LFSR data. The results are shown in Figure \ref{fig_misalign}. Both $V_{RECT}$ and the backscatter amplitude decrease smoothly with increasing lateral offset, consistent with the rolloff of the acoustic beam profile. The ASK communication protocol operates reliably within $\pm$2 mm of lateral offset, beyond which $V_{RECT}$ falls below the minimum operating threshold of 1.4V. As shown in Figure \ref{fig_misalign}(b), decoding LFSR data shows that no more than one level error was made in 6334 8-bit samples, demonstrating reliable communication within the $\pm$2 mm misalignment range; this indicates that the system has sufficient SNR margin to accommodate misalignment-induced amplitude degradation. As shown in Figure \ref{fig_misalign}(a), the $\pm$2 mm tolerance is primarily determined by the beam width of the transducer used in this work.

\section{Conclusion}

\subsection{Summary and Comparison}

A comparison to prior wireless, chip-based multi-implant systems is shown in Table II. The proposed system uses a 16-level ASK modulation scheme to transmit neural data over an ultrasound link, enabling a total data rate of 400 kb/s at a carrier frequency of 2 MHz. Using a TDMA protocol to communicate with multiple implants, the proposed system also demonstrates the highest number of simultaneously-recording US-powered implants to date. DustNet can operate at depths in excess of 90mm, which is \>15x deeper than comparable RF- or NIR-powered neural implants. Furthermore, the 16-level ASK modulation scheme enables highly efficient use of the channel bandwidth; it increases the data rate by a factor of 4 compared to equivalent OOK modulation schemes and enables a \>5x increase in the amount of information transmitted per US carrier cycle compared to existing analog AM modulation schemes for US implants \cite{Ghanbari_JSSC19}.

\subsection{Practical Considerations for In Vivo Use}

While this work validates the DustNet architecture in a controlled benchtop setup, transitioning to a biological environment introduces numerous physical non-idealities. The DustNet system addresses these challenges through the flexibility of the TDMA protocol and by placing the signal processing burden on the external interrogator, enabling adaptive protocol adjustment and active signal processing during operation to ensure high-fidelity data transmission.

In realistic physical conditions, angular or lateral misalignment results in increased channel attenuation, reducing the amplitude of the backscattered signal. DustNet mitigates misalignment-induced amplitude degradation through the interrogator’s adaptive k-means clustering algorithm, which determines optimal threshold values to decode ASK-modulated data based on the received signal distribution. The I-DAC backscatter modulator provides an additional 5.3dB of SNR margin compared to passive impedance modulation, further providing robustness to misalignment. Lateral misalignment tolerance was characterized experimentally; DustNet is able to tolerate a $\pm2$ mm lateral misalignment between the transducer and piezo before $V_{RECT}$ drops below the minimum operating voltage while making at most one level error in 50672 transmitted bits. The backscatter amplitude decreases smoothly within this range, and adaptive k-means thresholding compensates for the resulting amplitude reduction. The $\pm2$ mm tolerance is primarily limited by the beam width; a larger aperture transducer would provide a wider beam at depth and would therefore provide a larger misalignment tolerance for both power delivery and communication. Depth misalignment can be handled by adjusting the width of Uplink Mode pulses on a per-implant basis. The interrogator can calculate the ToF to each implant by measuring the time between transmission of an Uplink Mode pulse and reception of the modulated backscatter; it can then set the width of each TDMA slot to be twice the corresponding ToF.

\begin{table}[t]
\centering
\caption{Comparison of Wireless Neural Implant Systems}
\label{table02}
\renewcommand{\arraystretch}{1.5}
\setlength\tabcolsep{1.5pt}
{\scriptsize
\begin{tabular}{m{0.085\textwidth}<{\centering}|m{0.068\textwidth}<{\centering}|m{0.052\textwidth}<{\centering}|m{0.050\textwidth}<{\centering}|m{0.050\textwidth}<{\centering}|m{0.050\textwidth}<{\centering}|m{0.060\textwidth}<{\centering\arraybackslash}}
\hlineB{2}
& \textbf{This work} & \cite{Ghanbari_JSSC19} & \cite{lee_neural_2021} & \cite{Atzeni20} & \cite{Yu2020} & \cite{So2024} \\
\hline
Wireless Link & \textbf{US} & US & RF & RF / NIR & ME / Inductive & Ind. + US (Relay) \\
\hline
Max Depth [mm] & \textbf{90} & 50 & 5 & 2 & 60 & 22 \\
\hline
Data Uplink Protocol & \textbf{ASK16} & Analog AM & BPSK & LSK (RF) & LSK (Ind.) & 16-QAM MIMO \\ [1.5pt]
\hline
Downlink Protocol & \textbf{ASK2} & N/A & ASK-PWM & PWM (Opt.) & TDM (ME) & ASK-PPM (Ind.) \\ [1.5pt]
\hline
Multiple Access Protocol & \textbf{TDMA} & CDMA & TDMA & CDMA & FDMA & MIMO \\
\hline
Max \# of Implants & \textbf{8} & 2 & 770 & N/A & 15 & 1 \\
\hline
Measured \# of Implants & \textbf{4} & 2 & 48 & 4 & 15 (DL) / 1 (UL) & 1 \\ [2pt]
\hline
Carrier Freq. [MHz] & \textbf{2} & 1.78 & 915 & 3000 & 0.31 (DL) / 31 (UL) & 3 (US) / 40.68 (Ind.) \\
\hline
Meas. Data Rate (kb/s)* & \textbf{200} & 70 & 10000 & 0.000312 & 40 & 12000** \\
\hline
Process [nm] & \textbf{28} & 65 & 65 & 180 & 180 & 180 HV \\ [2pt]
\hline
IC Area [mm\textsuperscript{2}] & \textbf{0.43} & 0.25 & 0.42 & 0.07 & 0.99 & 9 \\ [2pt]
\hline
IC Power [µW] & \textbf{7} & 38 & 30 & 0.57 & 11 & N.R.$^\dagger$ \\ [2pt]
\hline
BER & \textbf{1.9E-5 @ 90mm} & N/A & 5E-3 @ 5mm & 1E-6 @ 2mm & 5.5E-5 @ 20mm & 1.1E-3 @ 22mm \\ [2pt]
\hlineB{2}
\end{tabular}
}
\begin{minipage}{\columnwidth}
{\scriptsize
\textsuperscript{*}DustNet supports up to 400 kb/s total uplink data rate.\\
\textsuperscript{$\dagger$}Relay architecture; total implant power not reported. Implant Tx power: 875\,µW.
\textsuperscript{**}Using multiple transducers for transmit and receive.
}
\\
\end{minipage}
\end{table}

The DustNet architecture also builds in robustness to physiological motion. The interrogator would be mounted on the skin over the target nerve, so the external transducer and implants move in tandem during limb movements; localized depth variations caused by muscle flexion or cardiac activity are constrained to a few millimeters. Given that the speed of sound in tissue ($c\approx 1540$ m/s), the round-trip ToF changes by only 1.3µs per mm of distance variation; this is less than 3 US carrier cycles at 2MHz and is negligible compared to the length of the TDMA timeslots. DustNet's demodulation algorithm is also robust to motion artifacts; because the ultrasound pulse duration is significantly shorter than the timescale of biological motion, motion artifacts affect the entire data packet equally and appear as common-mode noise that is rejected by the differential envelope decoding algorithm.

Multipath reflections from deeper bone structures produce a deterministic background echo that could be characterized and removed in post-processing if necessary. The external interrogator can transmit a calibration pulse prior to data transmission to characterize this background echo and subtract it from subsequent received signals in the digital domain; this does not require additional hardware or power overhead on the implant. Scaling the network beyond the four physically instantiated implants is supported by the dynamic range of the external interrogator, which is sufficient to handle the additional passive background reflections contributed by non-modulating implants in each TDMA frame without degrading the per-implant SNR.

\section*{Acknowledgments} 
The authors thank the National Science Foundation Graduate Research Fellowship, the Weill Neurohub Fellowship, and the sponsors of the Berkeley Wireless Research Center. The authors also thank TSMC for sponsoring chip fabrication.

\enlargethispage{-0.5in}

 


\ifCLASSOPTIONcaptionsoff
  \newpage
\fi



%

{\bibliographystyle{IEEEtran}
\bibliography{reference}

@article{lee_implantable_2018,
	title = {An {Implantable} {Peripheral} {Nerve} {Recording} and {Stimulation} {System} for {Experiments} on {Freely} {Moving} {Animal} {Subjects}},
	volume = {8},
	copyright = {2018 The Author(s)},
	issn = {2045-2322},
	doi = {10.1038/s41598-018-24465-1},
	language = {en},
	number = {1720},
	urldate = {2025-02-04},
	journal = {Scientific Reports},
	author = {Lee, Byunghun and Koripalli, Mukhesh K. and Jia, Yaoyao and Acosta, Joshua and Sendi, M. S. E. and Choi, Yoonsu and Ghovanloo, Maysam},
	month = apr,
	year = {2018},
	note = {Publisher: Nature Publishing Group},
	pages = {6115},
}

@article{rey_past_2015,
  title={Past, present and future of spike sorting techniques},
  author={Rey, Hernan Gonzalo and Pedreira, Carlos and Quiroga, Rodrigo Quian},
  journal={Brain research bulletin},
  volume={119},
  pages={106--117},
  year={2015},
  publisher={Elsevier}
}

@inproceedings{alamouti_high_2020,
	title = {High Throughput Ultrasonic Multi-implant Readout Using a Machine-Learning Assisted {CDMA} Receiver},
	doi = {10.1109/EMBC44109.2020.9176480},
	eventtitle = {2020 42nd Annual International Conference of the {IEEE} Engineering in Medicine \& Biology Society ({EMBC})},
	pages = {3289--3292},
	booktitle = {2020 42nd Annual International Conference of the {IEEE} Engineering in Medicine \& Biology Society ({EMBC})},
	author = {Alamouti, Sina Faraji and Ghanbari, Mohammad Meraj and Ersumo, Nathan Tessema and Muller, Rikky},
	urldate = {2025-02-05},
	date = {2020-07},
	note = {{ISSN}: 2694-0604},
}

@INPROCEEDINGS{So2024,
  author={So, Ernest and Arbabian, Amin},
  booktitle={2024 IEEE International Solid-State Circuits Conference (ISSCC)}, 
  title={6.1 12Mb/s 4×4 Ultrasound MIMO Relay with Wireless Power and Communication for Neural Interfaces}, 
  year={2024},
  volume={67},
  number={},
  pages={100-102},
  doi={10.1109/ISSCC49657.2024.10454377}}

@article{lee_neural_2021,
  title={Neural recording and stimulation using wireless networks of microimplants},
  author={Lee, Jihun and Leung, Vincent and Lee, Ah-Hyoung and Huang, Jiannan and Asbeck, Peter and Mercier, Patrick P and Shellhammer, Stephen and Larson, Lawrence and Laiwalla, Farah and Nurmikko, Arto},
  journal={Nature Electronics},
  volume={4},
  number={8},
  pages={604--614},
  year={2021},
  publisher={Nature Publishing Group UK London}
}

@article{abbott1999,
  title={Rationale and derivation of MI and TI—a review},
  author={Abbott, John G},
  journal={Ultrasound in medicine \& biology},
  volume={25},
  number={3},
  pages={431--441},
  year={1999},
  publisher={Elsevier}
}

@inproceedings{rabbani2021,
  title={Towards an implantable fluorescence image sensor for real-time monitoring of immune response in cancer therapy},
  author={Rabbani, Rozhan and Najafiaghdam, Hossein and Ghanbari, Mohammad Meraj and Papageorgiou, Efthymios P and Zhao, Biqi and Roschelle, Micah and Stojanovic, Vladimir and Muller, Rikky and Anwar, Mekhail},
  booktitle={2021 43rd Annual International Conference of the IEEE Engineering in Medicine \& Biology Society (EMBC)},
  pages={7399--7403},
  year={2021},
  organization={IEEE}
}

@ARTICLE{Ghanbari_JSSC19,  author={Ghanbari, Mohammad Meraj and Piech, David K. and Shen, Konlin and Faraji Alamouti, Sina and Yalcin, Cem and Johnson, Benjamin C. and Carmena, Jose M. and Maharbiz, Michel M. and Muller, Rikky},  journal={IEEE J. Solid-State Circuits},   title={A Sub-mm3 Ultrasonic Free-Floating Implant for Multi-Mote Neural Recording},   year={2019},  volume={54},  number={11},  pages={3017-3030},  doi={10.1109/JSSC.2019.2936303}}

@INPROCEEDINGS{Yu2020,
  author={Yu, Zhanghao and Wang, Wei and Chen, Joshua C. and Chen, Zhiyu and He, Yan and Singer, Amanda and Robinson, Jacob T. and Yang, Kaiyuan},
  booktitle={2022 IEEE Radio Frequency Integrated Circuits Symposium (RFIC)}, 
  title={A Wireless Network of 8.8-mm3 Bio-Implants Featuring Adaptive Magnetoelectric Power and Multi-Access Bidirectional Telemetry}, 
  year={2022},
  volume={},
  number={},
  pages={47-50},
  doi={10.1109/RFIC54546.2022.9863077}}

@inproceedings{Chang19,
  title={Multi-access networking with wireless ultrasound-powered implants},
  author={Chang, Ting Chia and Wang, Max and Arbabian, Amin},
  booktitle={2019 IEEE Biomedical Circuits and Systems Conference (BioCAS)},
  pages={1--4},
  year={2019},
  organization={IEEE}
}

@inproceedings{Chang17,
  title={27.7 A 30.5 mm 3 fully packaged implantable device with duplex ultrasonic data and power links achieving 95kb/s with< 10- 4 BER at 8.5 cm depth},
  author={Chang, Ting Chia and Wang, Max L and Charthad, Jayant and Weber, Marcus J and Arbabian, Amin},
  booktitle={2017 IEEE International Solid-State Circuits Conference (ISSCC)},
  pages={460--461},
  year={2017},
  organization={IEEE}
}

@ARTICLE{Roschelle_JSSC,
  author={Roschelle, Micah and Rabbani, Rozhan and Gweon, Surin and Kumar, Rohan and Vercruysse, Alec and Woo Cho, Nam and Spitzer, Matthew H. and Niknejad, Ali M. and Stojanović, Vladimir M. and Anwar, Mekhail},
  journal={IEEE Journal of Solid-State Circuits}, 
  title={A Wireless, Multicolor Fluorescence Image Sensor Implant for Real-Time Monitoring in Cancer Therapy}, 
  year={2024},
  volume={59},
  number={11},
  pages={3580-3598},
  doi={10.1109/JSSC.2024.3435736}}

@INPROCEEDINGS{CLee_ISSCC25,
  author={Lee, Changuk and Pinkenburg, Jade and Ghanbari, Mohammad Meraj and Yalcin, Cem and Montalban, Miguel and Muller, Rikky},
  booktitle={2025 IEEE International Solid-State Circuits Conference (ISSCC)}, 
  title={35.8 DustNet: A Network of Time-Division Multiplexed Ultrasonic Implants with 16-Level ASK Backscatter Modulation}, 
  year={2025},
  volume={68},
  number={},
  pages={582-584},
  doi={10.1109/ISSCC49661.2025.10904679}}

@article{Seo16,
  title={Wireless recording in the peripheral nervous system with ultrasonic neural dust},
  author={Seo, Dongjin and Neely, Ryan M and Shen, Konlin and Singhal, Utkarsh and Alon, Elad and Rabaey, Jan M and Carmena, Jose M and Maharbiz, Michel M},
  journal={Neuron},
  volume={91},
  number={3},
  pages={529--539},
  year={2016},
  publisher={Elsevier}
}

@article{Ghanbari_TBIOCAS20,
  title={Optimizing volumetric efficiency and backscatter communication in biosensing ultrasonic implants},
  author={Ghanbari, Mohammad Meraj and Muller, Rikky},
  journal={IEEE Transactions on Biomedical Circuits and Systems},
  volume={14},
  number={6},
  pages={1381--1392},
  year={2020},
  publisher={IEEE}
}

@inproceedings{Ozilgen17,
  title={Ultrasonic thermal dust: A method to monitor deep tissue temperature profiles},
  author={Ozilgen, B Arda and Maharbiz, Michel M},
  booktitle={2017 39th Annual International Conference of the IEEE Engineering in Medicine and Biology Society (EMBC)},
  pages={865--868},
  year={2017},
  organization={IEEE}
}

@inproceedings{Sonmezoglu20,
  title={34.4 A 4.5 mm 3 deep-tissue ultrasonic implantable luminescence oxygen sensor},
  author={Sonmezoglu, Soner and Maharbiz, Michel M},
  booktitle={2020 IEEE International Solid-State Circuits Conference-(ISSCC)},
  pages={454--456},
  year={2020},
  organization={IEEE}
}

@article{Vu2020,
  title={A regenerative peripheral nerve interface allows real-time control of an artificial hand in upper limb amputees},
  author={Vu, Philip P and Vaskov, Alex K and Irwin, Zachary T and Henning, Phillip T and Lueders, Daniel R and Laidlaw, Ann T and Davis, Alicia J and Nu, Chrono S and Gates, Deanna H and Gillespie, R Brent and Kemp, Stephen W P and Kung, Theodore A and Chestek, Cynthia A and Cederna, Paul S},
  journal={Science translational medicine},
  volume={12},
  number={533},
  pages={eaay2857},
  year={2020},
  publisher={American Association for the Advancement of Science}
}

@article{flint2012local,
  title={Local field potentials allow accurate decoding of muscle activity},
  author={Flint, Robert D and Ethier, Christian and Oby, Emily R and Miller, Lee E and Slutzky, Marc W},
  journal={Journal of neurophysiology},
  volume={108},
  number={1},
  pages={18--24},
  year={2012},
  publisher={American Physiological Society Bethesda, MD}
}

@article{stavisky2015high,
  title={A high performing brain--machine interface driven by low-frequency local field potentials alone and together with spikes},
  author={Stavisky, Sergey D and Kao, Jonathan C and Nuyujukian, Paul and Ryu, Stephen I and Shenoy, Krishna V},
  journal={Journal of neural engineering},
  volume={12},
  number={3},
  pages={036009},
  year={2015},
  publisher={IOP Publishing}
}

@article{Vancara23,
  title={Bringing sensation to prosthetic hands—chronic assessment of implanted thin-film electrodes in humans},
  author={{\v{C}}van{\v{c}}ara, Paul and Valle, Giacomo and M{\"u}ller, Matthias and Bartels, Inga and Guiho, Thomas and Hiairrassary, Arthur and Petrini, Francesco and Raspopovic, Stanisa and Strauss, Ivo and Granata, Giuseppe and Fernandez, Eduardo and Rossini, Paolo M and Barbaro, Massimo and Yoshida, Ken and Jensen, Winnie and Divoux, Jean-Louis and Guiraud, David and Micera, Silvestro and Steiglitz, Thomas},
  journal={Npj Flexible Electronics},
  volume={7},
  number={1},
  pages={51},
  year={2023},
  publisher={Nature Publishing Group UK London}
}

@ARTICLE{Paidimarri16,
  author={Paidimarri, Arun and Griffith, Danielle and Wang, Alice and Burra, Gangadhar and Chandrakasan, Anantha P.},
  journal={IEEE Journal of Solid-State Circuits}, 
  title={An RC Oscillator With Comparator Offset Cancellation}, 
  year={2016},
  volume={51},
  number={8},
  pages={1866-1877},
  doi={10.1109/JSSC.2016.2559508}}

@INPROCEEDINGS{Atzeni20,
  author={Atzeni, Gabriele and Lim, Jongyup and Liao, Jiawei and Novello, Alessandro and Lee, Jungho and Moon, Eunseong and Barrow, Michael and Letner, Joseph and Costello, Joseph and Nason, Samuel R. and Patel, Paras R. and Patil, Parag G. and Kim, Hun-Seok and Chestek, Cynthia A. and Phillips, Jamie and Blaauw, David and Jang, Taekwang},
  booktitle={2022 IEEE Symposium on VLSI Technology and Circuits (VLSI Technology and Circuits)}, 
  title={A 260×274 μm2 572 nW Neural Recording Micromote Using Near-Infrared Power Transfer and an RF Data Uplink}, 
  year={2022},
  volume={},
  number={},
  pages={64-65},
  doi={10.1109/VLSITechnologyandCir46769.2022.9830516}}
}
%







\end{document}